\documentclass[10pt,journal,compsoc]{IEEEtran}

\usepackage[nocompress]{cite}
\usepackage{hyperref}
\usepackage{xspace}
\usepackage{xcolor}
\usepackage{amsmath}
\usepackage{amssymb}
\usepackage{graphicx}

\newcommand{\K}{$\textrm{k}\mathsf{^2}\textrm{-tree}$\xspace}
\newcommand{\Ks}{$\textrm{k}\mathsf{^2}\textrm{-trees}$\xspace}

\newcommand{\subject}{{subject}\xspace}
\newcommand{\predicate}{{predicate}\xspace}
\newcommand{\object}{{object}\xspace}
\newcommand{\subjects}{{subjects}\xspace}
\newcommand{\predicates}{{predicates}\xspace}
\newcommand{\objects}{{objects}\xspace}

\newcommand{\no}[1]{}

\newcommand{\csa}{{\it CSA}\xspace}
\newcommand{\icsa}{{\it iCSA}\xspace}
\newcommand{\rdfcsa}{{\it RDFCSA}\xspace}

\newcommand{\kdostriples}{{K2Triples}\xspace}
\newcommand{\kdostriplesplus}{{K2Triples+}\xspace}
\newcommand{\trie}{{\em trie}\xspace}
\newcommand{\trietwo}{{\em trie-2tp}\xspace}
\newcommand{\trieb}{{\em trie-3t}\xspace}

\begin{document}
%
\title{Space/time-efficient RDF stores based on circular suffix sorting}
%
%
%
%

\author{Nieves R. Brisaboa,
        Ana Cerdeira-Pena,
        Guillermo de Bernardo,
        Antonio Fari\~na,
        and~Gonzalo Navarro
\IEEEcompsocitemizethanks{\IEEEcompsocthanksitem G. Navarro is affiliated to DCC, University of Chile and Millenium Institute for Foundational Research on Data (IMFD), Santiago, Chile.\protect\\
\IEEEcompsocthanksitem N. R. Brisaboa, A. Cerdeira-Pena, G. de Bernardo and A. Fari\~na are affiliated to Universidade da Coru\~na and Centro de Investigaci\'on CITIC, A Coru\~na, Spain.}
\thanks{An early partial version of this article appeared in {\em Proc
SPIRE'15} \cite{spire}.}}

\IEEEtitleabstractindextext{%
\begin{abstract}
In recent years, RDF has gained popularity as a format for the standardized
publication and exchange of information in the Web of Data. In this paper we
introduce \rdfcsa, a data structure that is able to self-index an RDF dataset in
small space and supports efficient querying.
\rdfcsa regards the triples of the RDF store as short circular strings and applies suffix sorting on those strings, so that triple-pattern queries reduce to prefix searching on the string set. The RDF store is then represented compactly using a {\em Compressed Suffix Array (CSA)}, a proved technology in text indexing that efficiently supports prefix searches.

Our experiments show that \rdfcsa provides a compact RDF representation, using less than 60\% of the space required by the raw data, and yields fast and consistent query times when answering triple-pattern queries (a few microseconds per result). We also support join queries, a key component of most SPARQL queries. \rdfcsa is shown to provide an excellent space/time tradeoff, typically using much less space than alternatives that compete in time.
\end{abstract}

}


\maketitle

\IEEEdisplaynontitleabstractindextext

%
\IEEEpeerreviewmaketitle

\section{Introduction}

Since the advent of the World Wide Web a few decades ago, the volume of
publicly available data has been increasing at a fast pace and has become an invaluable repository of information at global scale, scattered along a large
number of repositories from several sources.
Since it was originally designed for direct human use, most of such information is stored in the form of unstructured Web pages and hyperlinks between
them, which limits our ability to automatically access and process it.
The Web of Data is an effort to provide a formal structure on the data, so that
it can be published and processed in automatic form. The Web of Data builds on 
top of the concepts of the Semantic Web~\cite{Berners2001}.

The Resource Description Framework (RDF)~\cite{RDFw3crecomendation,RDFw3crecomendation2} is a
W3C recommendation designed to publish and share information in the
Web of Data. It is based on a simple labeled-graph-like conceptual structure, but it
does not enforce a specific storage format. This graph is usually regarded,
for most practical purposes, as a collection of triples, or 3-tuples (source, label, target), that
represent the edges in the graph. Going further in the
standardization effort, a specific query language called SPARQL has been
defined~\cite{SPARQL:2008} to query RDF collections. SPARQL is based on the
concept of triple pattern, a tuple that may contain some unbound elements and
that is matched against all the triples in the RDF dataset. Building on this basic selection query, SPARQL enables matching of more complex subgraphs by means of joins, which connect triples that share some component. 


The ability of RDF to provide a simple format to publish information has led
to its rise in popularity in recent years. The lack of an enforced physical
representation format has also led to the emergence of many different
solutions to efficiently store the RDF data. These solutions, generally
called RDF stores or triple stores, aim at providing efficient storage and
querying of the RDF dataset. Some RDF stores rely on adapting existing ideas from relational or graph databases~\cite{Sakr:2010}. Tools such as Virtuoso~\cite{virtuoso} and Blazegraph~\cite{Blazegraph}, work as fully-functional RDF stores and provide a wide range of query capabilities. Other solutions are based on custom techniques devised specifically for RDF or adapted from other areas. Some examples of these tools include RDF-3X~\cite{Neumann:2010}, Tentris~\cite{tentris}, BITMAT~\cite{Atre:2010}, HEXASTORE~\cite{Weiss:2008}, WaterFowl~\cite{Cureetal.ESWC:2014}, or HDT~\cite{FernaNdez:2013:HDT}.

The main issue for modern RDF stores, as the number and size of RDF
datasets increases, is the scalability of the solutions~\cite{Jing:2009}.
New approaches have been proposed to tackle this problem. Most
solutions based on databases or custom indexes rely on caching to
maintain good query performance even if the full dataset is too large to fit in
main memory. New proposals of distributed stores~\cite{hadooprdf,dream} provide
a framework to store and query in a clustered environment, thus facilitating
scalability. Finally, a number of solutions aim at achieving very efficient
compression so that even large datasets can be efficiently stored and queried in
main memory in regular machines, based on compact data
structures; \kdostriples~\cite{AGBFMPNkais14} and permuted trie
indexes~\cite{Venturini} are examples of 
proposals that work in this way.
Both \kdostriples and permuted trie indexes assume that RDF triples are composed of numeric identifiers, so they rely on an external compact dictionary to map RDF strings to identifiers \cite{Martinez-Prieto:2012,migueldiccionarios}.

In this paper we introduce \rdfcsa, a solution for the
compact representation of RDF data that aims at combining good compression with consistently good query performance. \rdfcsa is based on the
compressed suffix array, or \csa~\cite{Sad03}, a data structure originally
devised for text indexing that is able to store a set of sequences in compressed
space and efficiently supports prefix searches. We modify the \csa
to regard the triples of the RDF dataset as short circular strings. All the triple-pattern queries can then be transformed into appropriate prefix searches, which are efficiently solved with the \csa. Join queries can also be implemented by exploiting the query capabilities of the \csa. We further engineer the \csa to optimize its performance in this scenario. 

We test our proposal against a variety of state-of-the-art solutions. Our
experimental results show that our solution provides an excellent space/time
tradeoff with respect to other solutions: \kdostriples obtains better
compression but is significantly slower than \rdfcsa, whereas permuted trie indexes are uniformly faster only when using significantly more space. Additionally, our results show that, thanks to its uniform treatment of all triple patterns, the query times of \rdfcsa
are very consistent and predictable. We also perform comparisons with other popular
representations, including HDT, Virtuoso, Blazegraph, MonetDB, RDF-3X, and Tentris; all of these are shown to be far from
competitive with \rdfcsa, being in most cases several times larger and/or several orders of magnitude slower.

The rest of this paper is organized as follows:
Section~\ref{sec:stateoftheart} provides some additional details about RDF,
as well as some of the relevant state-of-the-art alternatives, and explains
the elements of the \csa data structure, necessary to understand our solution.
Section~\ref{sec:RDFCSA} describes the \rdfcsa data structure, and the basic
algorithms for simple and advanced queries. 
Section~\ref{sec:experiments} details the experimental evaluation performed.
Finally, Section~\ref{sec:conclusions} presents the main conclusions of this
work and outlines future work.

\section{Previous concepts and related work}
\label{sec:stateoftheart}

\subsection{RDF, triple patterns, and SPARQL}
\label{sec:basic_rdf}

The RDF data model is based on a graph-like representation of the data, where
information about a set of entities is conceptually stored using labeled arcs in
a directed graph. Given an entity ({\em subject}), that is associated with a
node, each of its properties will be represented with an outgoing arc (labeled by a {\em
	predicate}), pointing to another node ({\em object}) that represents the
value of that property~\cite{RDFw3crecomendation}. An especially useful way of
seeing this graph, that is also proposed in the definition of the format, is as
a collection of {\em triples}: we consider that an RDF dataset is a set
$\mathcal{R}$ of triples $(s,p,o)$ (i.e. subject, predicate, object), where each triple represents an arc of the
graph.

\begin{figure}[htbp]
	\centering
	\includegraphics[width=1.0\linewidth]{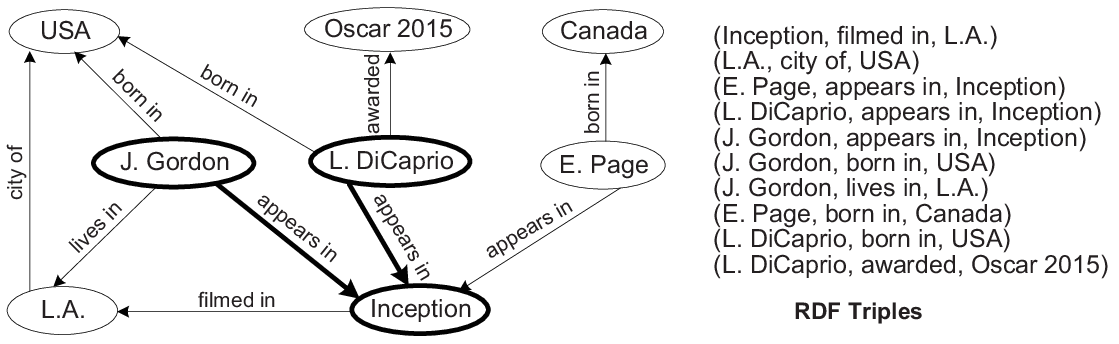}
	\caption{Example of RDF graph and its representation as a set of triples.}
	\label{fig:rdf}
\end{figure}

Figure~\ref{fig:rdf} displays an example of an RDF dataset, represented as a graph
or as a set of string triples. Each triple represents an edge of the
graph, storing the source node as the subject, the label as the predicate, and the
target node as the object. Note that we are using simple strings to denote
subjects, predicates, and objects. Yet in RDF, subjects and predicates must
always be identified with URIs, whereas objects may be either URIs or literal
values (we are omitting some other artifacts of RDF, such as blank nodes, since they are not relevant for this work; for our purposes, it suffices to regard each component as any kind of string).

RDF collections can be queried using the SPARQL query language. SPARQL is a complex language with many features, but at its core are triple patterns. 
A triple pattern is a tuple
$(subject,predicate,object)$ where each of its elements may be either {\em
	bound} or {\em unbound}. For instance, the pattern $(s,p,o)$, where all three elements
are bound, asks whether subject $s$ has a predicate (or ``property'') $p$ with value $o$; the pattern $(s,p,?o)$, where the object is unbound, asks for the objects to which subject $s$ is associated via predicate $p$; the
pattern $(s,?p,?o)$, where both predicate and object are unbound, asks for all
the pairs $(p,o)$ corresponding to the properties of subject $s$.

SPARQL queries can express more complex conditions using a combination of triple patterns. In this kind of queries, the triple patterns are usually combined using {\em join variables}, that is, elements of
different triple patterns that must take the same value. For instance, the
simple join operation $(s_1,p_1,?x) \bowtie (s_2, p_2, ?x)$ (where $?x$ is the join variable) asks for all the
objects that are associated to $s_1$ by property $p_1$ and to $s_2$ by property $p_2$. For instance, to
know the names of the movies where both {\em L.
	DiCaprio} and {\em J. Gordon} appeared in, we could ask for $(\mathit{L.~
	DiCaprio}, \mathit{appears~in}, ?x) \bowtie (\mathit{J.~Gordon}, \mathit{appears~in},
?x)$, and it would return the movie {\em Inception}, as highlighted in
Figure~\ref{fig:rdf}.

A wide variety of join operations can be performed depending on the bound and
unbound elements in each individual pattern and also on the position of the
join variables. For instance, the previous example $(s_1,p_1,?x) \bowtie (s_2,
s_2, ?x)$ is an {\em object-object} join, because the join variable plays the role of object in both triples; the equivalent {\em subject-object} 
and {\em subject-subject} joins would be $(s_1,p_1,?x) \bowtie (?x,
p_2, o_2)$ and $(?x,p_1,o_1) \bowtie (?x,
p_2, o_2)$, respectively. Additionally, we may also categorize joins
according to the unbound elements that appear in one or both of the patterns
(e.g.
$(?s_1,?p_1,?x) \bowtie (?s_2, p_2, ?x)$, and $(?s_1,p_1,?x) \bowtie (?s_2,
p_2, ?x)$ are different types of joins because they differ in the number of
unbound elements). For example, $(?x, \mathit{appears~in}, ?y) \bowtie (?x, \mathit{lives~in}, ?z) \bowtie (?y, \mathit{filmed~in}, ?z)$ looks for actors appearing in a movie filmed in the city where they live. This yields the binding $x=\mathit{J.~Gordon}$, $y=\mathit{Inception}$, $z=\mathit{L.A.}$ in Figure~\ref{fig:rdf}.

A set of triple patterns such as the examples above, with any number of triples and join variables, is usually denoted as a basic graph pattern (BGPs). This is the key component that appears in almost all SPARQL queries. A basic graph pattern is a generic set of triple patterns, and may involve any number of join variables, even though most real-world queries follow typical patterns. 
In this paper, we focus on the performance for the execution of simple binary join queries, involving just two triple patterns. As we will see, the join techniques used in this paper can be easily extended for joins of any number of patterns. However, for larger BGPs the execution order of the joins and the selection of join technique in each case become more challenging. 


\subsection{RDF stores}

As stated before, multiple solutions have been developed to efficiently store
and query RDF datasets. The most popular RDF stores are fully functional systems that provide not only storage and query capabilities, but also update mechanisms and 
integrated SPARQL query endpoints. Virtuoso~\cite{virtuoso} and Blazegraph~\cite{Blazegraph} are two representative examples of database solutions with all of these functionalities. 

In addition to these popular solutions, many other representations have been proposed with varying capabilities and focus, regarding their query support, update capabilities, etc. In this paper, we focus on lower-level solutions, that tackle the compact storage of the underlying data by means of compact data structures, and aim at providing fast response times for triple pattern and join queries, without attempting to support all the capabilities of SPARQL and the features of a full database engine. Particularly, in this section we introduce several relevant RDF stores that are based on different compact data structures or indexing solutions. Among them, HDT and \kdostriples are of special interest to understand our work, as we share some ideas with them.

\subsubsection{HDT and dictionary encoding}

HDT~\cite{rdfhdt2,FernaNdez:2013:HDT} is a solution for RDF storage and
querying.
It was originally devised as a serialization format to take advantage of the redundancy that is usual in
RDF datasets, but it has gained popularity~\cite{FernaNdez:2013:HDT} thanks to its ability
to achieve a relatively good compression, and its support for basic SPARQL 
queries~\cite{hdtfoq}.
One key idea in HDT is the separation of the RDF dataset in three main
components: \emph{H}eader, \emph{D}ictionary, and \emph{T}riples. The Header
component simply stores metadata, and is not relevant for this paper. The
Dictionary stores the different strings appearing in the original RDF dataset,
and is in charge of assigning a numeric identifier to each string and providing
a bijective string-to-id translation. Finally, the Triples component stores the
triples themselves, where each triple is a tuple with three numeric
identifiers. This is relevant to our work since \rdfcsa essentially solves the
storage of the triples, and is compatible with the dictionary solutions in HDT,
so it could be used to replace its Triples component.

HDT defines the decomposition format and provides basic implementations for the
dictionary and the triples. Solutions for the dictionary are based on
sorting and removing redundancy from the collection of strings, although
further work has been pursued by the
authors~\cite{Martinez-Prieto:2012,migueldiccionarios}. Basic solutions for the triples rely on sorted lists that store their elements.
Although originally designed for publication and exchange of RDF, HDT can also
be used to query the data by enhancing the basic structure with additional
indexes.

\begin{figure}[htb]
	\centering
	{\includegraphics[width=1.0\linewidth]{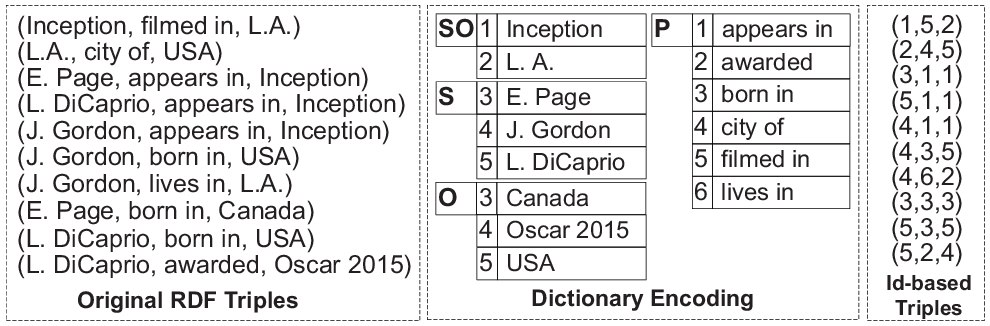}}
	\caption{Dictionary encoding used in HDT for the set of triples in
		Figure~\ref{fig:rdf}.}
	\label{fig:dict}
\end{figure}

Figure~\ref{fig:dict} displays the dictionary encoding used in HDT for the
set of triples from Figure~\ref{fig:rdf}. Strings are separated in four
different sets: a first set $SO$ contains strings that are both subjects and objects, and
then three other sets store subjects $S$, predicates $P$, and objects $O$. Each
set is sorted in lexicographic order, and correlative identifiers are assigned
to the elements of each set: entries in $SO$ and $P$ are numbered starting at 1,
and entries in $S$ and $O$ are numbered starting at $|SO|+1$. This is useful for
dictionary compression and guarantees that each subject, predicate, and object
has a unique identifier.

\subsubsection{\kdostriples}
\label{sec:basic_ktriples}

\kdostriples~\cite{AGBFMPNkais14} is a solution for the compact
representation of RDF triples. Like \rdfcsa, it only considers the
structural part of RDF, assuming that triples consist of integer
identifiers; also, like \rdfcsa, \kdostriples is compatible with the
dictionary scheme in HDT, and is focused on the efficient compression of the
triples.

The key idea in \kdostriples is the vertical partitioning~\cite{Abadi:2007} of
the data. Relying on the fact that the number of predicates (i.e., the
number of different properties) is usually very small in RDF datasets, vertical
partitioning separates the set of $(s,p,o)$ triples into one set per distinct predicate $p$, each containing the $(s,o)$ pairs connected by that predicate.
In \kdostriples, each set of pairs $(s,o)$ is regarded as a
binary relation and stored using a \K~\cite{BLNis13.2}. The \K not only
permits effectively compressing each binary relation, but its indexing capabilities are exploited to efficiently solve most queries in
\kdostriples by translating them into basic operations on the \Ks.

The authors have also proposed specific query algorithms to efficiently
answer queries involving joins of two triple patterns, as well as a
variation called \kdostriplesplus that improves the performance in queries with
unbound predicates. Those queries, which are usually the weak point in
techniques based on vertical partitioning, would require accessing all the \Ks
in the original \kdostriples, so the authors integrate all the binary relations and add additional indexes $SP$ and
$OP$ in order to reduce the number of structures that need to be accessed. This
drastically improves their performance at the cost of up to 30\% extra space.
Even with these additional indexes, \kdostriples variants are, to
the best of our knowledge, the most compact representations of RDF datasets with
efficient query support.

\subsubsection{Permuted trie index}

The permuted trie index is a recent RDF representation based
on the use of compressed tries~\cite{Venturini}. The index relies on the construction of
several permutations of the triples. In the basic
proposal, they use the permutations $SPO$,
$POS$, and $OSP$. Triple-pattern queries are
answered by accessing the appropriate structure depending on the fixed variables in the triple pattern.

The authors store each permutation as a 3-level trie, and propose several
compression techniques based on
Partitioned-Elias-Fano (PEF)~\cite{pef} compression, in order to obtain
different space/time tradeoffs. Their PEF-compressed tries show very good
performance in comparison with other state-of-the-art solutions.

In addition to their basic proposal, based on three indexes (which we refer to
as \emph{trie-3t}), they also propose solutions that aim at better compression
by removing one of the permutations from the index. The key idea of these variants is that, by removing
one of the indexes, queries that used the other two permutations are not
affected in performance, while some queries that used the removed permutation can
still be performed reasonably using the remaining ones. Among them, the best
choice~\cite[Sec.~4.1]{pef} is the variant that removes the permutation
$OSP$. We refer to it as \emph{trie-2tp}.

\subsection{Rank and select on bitmaps}

Bitmaps are the most fundamental components of compressed data structures. A bitmap $B[1,n]$ can be represented in plain form using $n$ bits of space, and then some relevant operations can be implemented on top of it by adding $o(n)$ extra bits.

The most basic operation of this kind is $rank_b(B,i)$, which counts the number of times bit $b$ appears in $B[1,i]$. This operation is easily computed in $\mathcal{O}(1)$ time with $o(n)$ extra bits~\cite{rank, Cla96}. The inverse operation, $select_b(B,j)$, finds the position of the $j$th occurrence of bit $b$ in $B$, and can also be computed in constant time using $o(n)$ additional bits \cite{Cla96,Mun96}.

In \rdfcsa, we only need $rank_1$ and $select_1$ operations, for which we build on a variant that requires $0.375n$ extra bits \cite{GGMN05}. We solve $rank_1$  using a two-level structure that, in the first level (superblocks), stores the cumulative values every {256} positions in an array using $(n/256)$ 32-bit integers, and in the second level (blocks), keeps the cumulative counters relative to the beginning of the corresponding superblock using $(n/32)$ 8-bit integers. We then compute $rank_1(B,i)$ by summing the counters at superblock $(i-1)/256$, and at block $(i-1)/32$, and finally scanning a 32-bit integer $u$ (the one covered by the corresponding block) to count the number of bits set up to position $i' = (i-1)~  \!\!\!\mod 32$. This last step can be solved in $\mathcal{O}(1)$ time using a popcount operation.
Instead, we used mask-and-shifting to set the bits $\geq i'$ from $u$ to zero, followed by four lookups to a $256$-byte table that indicates the number of bits set for any possible byte value. This yields $\mathcal{O}(1)$ time for $rank_1$.

For $select_1$, whose constant-time solution is not so practical, this variant \cite{GGMN05} binary searches the values sampled for $rank$ in the superblocks, then sequentially scans the counters of the blocks (up to $8$ accesses to block counters) to find the block that contains the $1$ we are looking for. Then, it scans the final 32-bit block using at most $4$ lookups into a 256-byte table, to locate the byte that contains that $1$. Finally, a lookup to a $256\!\times\!8$-byte table gives the position within the last byte of our $1$, completing $select_1$. Therefore, $select_1$ is solved in $\mathcal{O}(\log n)$ time, using essentially the same $rank$ structures. We later describe some improvement we make on top of this $select_1$ algorithm.

\subsection{Sadakane's Compressed Suffix Array}\label{sec:basic_csa}

The \emph{suffix array}~\cite{suffixArray} is a data structure widely used for
text indexing. Given a sequence $T[1,n]$, built over an alphabet
$\Sigma=[1,\sigma]$, its suffix array is an array $A[1,n]$ that contains a permutation
of the integers in $[1,n]$ such that $T[A[i],n] < T[A[i+1],n]$ for all $i$, in lexicographic order. The suffix array is  built by sorting all the suffixes $T[j,n]$
and storing in $A[i]$ the offset in the sequence $T$ of the
$i$th suffix in lexicographical order. Note that all the suffixes starting with the same string $\alpha$ are contiguous in $A$, and that any occurrence of $\alpha$ in $T$ is the prefix of a suffix of $T$ starting with $\alpha$. We can then efficiently search for all the occurrences of a pattern $\alpha[1,m]$ in $T$ by two binary searches on its suffix array $A$, requiring time $\mathcal{O}(m\log n)$, which locate the range $A[l,r]$ corresponding to all the positions where $\alpha$ occurs in $T$.

The original suffix array is useful for searching but requires a
significant amount of space, $n\log n$ bits, in addition to the original sequence.
Sadakane's Compressed Suffix Array, or \csa~\cite{Sad03}, provides a
compact representation that uses at most $n\log\sigma + \mathcal{O}(n\log\log\sigma)$ bits and replaces both $T$ and $A$, while still
efficiently supporting searches. 

The \csa is composed of several data structures. The most important
of them is a new permutation $\Psi[1,n]$~\cite{GV00}. For any $i$ in
$[1,n]$, assuming $A[i]=p$, $\Psi[i]$ stores the position $j$ in the suffix array that points
to the next position in the original sequence (i.e., $A[j]=A[i]+1=p+1$). A
special case arises when $A[i]=n$, where $\Psi[i]$ is set to $j$ such that $A[j]=1$. Concisely, $\Psi$
is defined as $\Psi[i] =A^{-1}[(A[i]~\!\!\bmod n)+1]$.

In addition to $\Psi$, a bitmap $D[1,n]$ contains a $1$ at the
positions in $A$ where the first symbol of the corresponding suffixes changes
(i.e., $D[i]=1$ iff $i=1$ or $T[A[i]] \not= T[A[i-1]]$). In order to know the
symbol in $T$ pointed by $A[i]$, we can count the number of $1$s in $D$ up to
position $i$, that is, $rank_1(D,i)$.

Using $\Psi$ and $D$ we can reproduce the same binary search of the suffix
array, without storing $T$ or $A$. The first symbol of the suffix pointed by
$A[i]$ can be computed as $rank_1(D,i)$. To extract the following symbols, we
iterate using $\Psi$: $\Psi[i]$ stores the position $i'$ in $A$ that points to
the next symbol of the text; therefore, we can extract
subsequent symbols as $rank_1(D,\Psi[i])$, $rank_1(D,\Psi[\Psi[i]])$, and so on.
Assuming that $rank$ operations in $D$ and accesses to $\Psi$ can be computed in
constant time, a binary search in the \csa still requires $\mathcal{O}(m\log n)$ time.
After computing the range $A[l,r]$ of the occurrences of $\alpha$, a forward text context for each can be extracted by  iterating with $\Psi$ in the same way. 

An uncompressed $\Psi$ array would still require the same space as $A$. However,
$\Psi$ can be partitioned into at most $\sigma$ increasing contiguous subsequences, which makes it highly compressible by encoding it differentially, i.e. by representing each $\Psi[i]$ as $\Psi[i]-\Psi[i-1]$.
A run of $t$ increasing values in $[1,n]$ can be represented in $t\log_2(n/t) + \mathcal{O}(t\log\log(n/t))$ using $\delta$-codes. Overall, $\Psi$ can be compressed to
space proportional to the zero-order empirical entropy of the original sequence,
or $nH_0(T)+\mathcal{O}(n\log H_0(T)) \le n\log\sigma + \mathcal{O}(n\log\log\sigma)$
bits~\cite{Sad03}. Further improvements, combining the $\delta$-codes with run-length encoding (RLE) for runs of consecutive differences equal to 1 (which tend to appear in $\Psi$), reduced this space even more and achieved compression
proportional to the higher-order entropy of $T$, $nH_k(T)$~\cite{NM07}. 

The \rdfcsa is based on the {\em integer-based CSA} (\icsa)%
\footnote{\url{http://vios.dc.fi.udc.es/indexing/wsi/}} 
\cite{FBNCPR12}. The \icsa is a variant optimized for large (integer-based) alphabets,
with some differences in implementation and compression techniques with
the original \csa. Particularly, in the \icsa the best compression is
achieved by using differential encoding of the consecutive $\Psi$ values, followed by mixing Huffman and run-length encoding of the resulting gaps. To provide
efficient access (in time $\mathcal{O}(t_\Psi)$) to $\Psi$, absolute $\Psi$ values are stored at positions $\Psi[1+k
\cdot t_{\Psi}], k \geq 0$.

Note that both the \csa and the \icsa include additional structures to support other text search functionalities. Particularly, they add samplings of $A$ and $A^{-1}$, to be able to find the position in $T$ of the occurrences of $\alpha$, or to extract arbitrary substrings. These additional data structures are not necessary in our \rdfcsa. 

\section{Our proposal: {\em RDFCSA}} \label{sec:RDFCSA}

The two compact approaches we reviewed in the previous section have issues to support all the possible combinations of triple patterns. \kdostriples and \kdostriplesplus are weaker when the predicate is unbound, whereas the permuted trie index favors the triple patterns where there is a trie starting with the bound elements. The key idea of \rdfcsa is that, if we regard the triples $(s,p,o)$ as circular strings (i.e., the $s$ follows the $o$ again), then for every possible triple pattern there is a rotation of $(s,p,o)$ where all the bound values precede all the unbound ones. Thus, if we index the triples as circular strings, every possible triple pattern can be reduced to a search for the circular strings that  start with some prefix. We use the \csa to simulate a set of circular strings corresponding to all the triples of the RDF dataset. This approach yields a uniform search approach that will translate into not only fast, but also consistent and predictable, query times.

We follow the convention of treating an RDF dataset as a set
$\mathcal{R}$ of triples $(s,p,o)$, where $s$, $p$, and $o$ are a
\subject, a \predicate, and an \object, respectively. Our
solution is designed to work with integer identifiers (ids) for each of them, so
it requires a separate dictionary to perform the translation between the
original string values and the corresponding integer ids. Particularly, we base our solution on the
same dictionary encoding proposed by HDT and also used by \kdostriples, which was
described in Section~\ref{sec:stateoftheart}.
Therefore, we assume a dictionary encoding in which subjects, predicates, and
objects are integers in contiguous ranges:
$s\in [1,n_s]$, $p\in [1,n_p]$, and $o\in [1,n_o]$ (note the overlapped identifiers in Figure~\ref{fig:dict}). While any other
dictionary encoding scheme could be used for our purposes without affecting
our implementation, we do take advantage of this
particular encoding to perform some optimizations in join queries.

Our \rdfcsa representation is a {\em self-index}, meaning that we can recover the triples from it, and thus it replaces the RDF store. As explained, it organizes the triples in a way that can be represented with a modified
\csa data structure that efficiently answers relevant queries in the domain. We first describe how the data structure is built from the set of triples,
and then how we efficiently support the relevant query
operations over our self-indexed representation of the triples.

\subsection{Data structure}

Given an input set $\mathcal{R}$ of $n$ triples, we sort them increasingly by
subject, then break ties using the \predicate and further break ties using the \object, to make up a sequence
$T_{sort}[1,n]$ of triples. Then, we transform this sequence of tuples into an
integer sequence of identifiers $T_{id}[1,3n]$, by placing the ids of the
three components of each entry $T_{sort}[i]$ at consecutive
positions $T_{id}[1+3(i-1)]$, $T_{id}[2+3(i-1)]$, and $T_{id}[3+3(i-1)]$. Hence, at the end of this step, $T_{id}[1,3n] = \langle s_1 ,p_1,o_1,$
$s_2 ,p_2,o_2,\dots, s_n ,p_n,o_n\rangle$ stores all the ids for the sorted triples.

Next, we transform the identifiers in order to obtain disjoint integer alphabets
$\Sigma_s$, $\Sigma_p$, and $\Sigma_o$ for the $n_s$ subjects, the $n_p$
predicates, and the $n_o$ objects. This can be performed just by
computing the displacements necessary for predicates and objects: we set an
array $ gaps[0,2] = [0,n_s,n_s+n_p]$ and convert sequence $T_{id}[1,3n]$ into
$T[1,3n]$, where $T[i] = T_{id}[i]+ gaps[(i-1) \bmod 3]$. After this
transformation, our sequence $T[1,3n]$ has an alphabet $\Sigma=[1,n_s+n_p+n_o]$,
where values in the range $[1,n_s]$ are reserved to subjects,
those in the range $[n_s+1,n_s+n_p]$ to predicates, and the remaining ones to objects. 

After the previous transformations, which can be trivially reversed to obtain
the original set $\mathcal{R}$ of triples, we build an \icsa on $T$.
However, some key changes have to be performed over the underlying suffix array
in order to efficiently answer queries. Those changes rely on specific properties of our
construction method. 

In particular, we take advantage of the following property of
the generated suffix array $A$: it contains three well-delimited sections
$A_s=A[1,n]$, $A_p=A[n+1,2n]$ and $A_o=A[2n+1,3n]$, corresponding respectively
to subjects, predicates, and objects. This is a direct consequence of our
construction method, which generates integer identifiers such that every \subject
is smaller than every \predicate, and this in turn is smaller than every
\object. This ordering means that, when sorting suffixes, entries corresponding
to subjects, predicates, and objects end up clustered in different sections.
Therefore, $A_s$ contains entries pointing to subjects in $T$, $A_p$
points to predicates, and $A_o$ points to objects. Accordingly, array $\Psi$ also
contains three separate ranges with special properties.
Recall that $\Psi[i]$ contains, for the position $p$ such that $A[i]=p$,
the position in $A$ that points to the next element $p+1$ in $T$. Due to the
division of $A$ into three sections, entries in $\Psi$ also point to
those delimited intervals, so each region of $\Psi$ contains values in
a different range: values of $\Psi[1,n]$ are in the range $[n+1,2n]$
(pointing to the range of \predicates); entries in $\Psi[n+1,2n]$ are in
the range $[2n+1,3n]$ (pointing to \objects); and entries in
$\Psi[2n+1,3n]$ are in the range $[1,n]$ (pointing to \subjects).

Since our sequence $T$ contains all the concatenated triples in $SPO$ order,
the symbol following an object will always be the subject of the next triple.
Therefore, if we are at position $i$ in the suffix array, such that $A[i]$
points to an object (i.e., $A[i]$ for $i \in [2n+1,3n]$, or $A[i]=3k$
for some $k$), when we iterate using $\Psi$ we reach a position $j$ such
that $A[j]$ points to the subject of the next triple. The original organization
of $\Psi$ was useful in the \csa to allow full extraction of the text.
In our case, however, we only need to extract individual triples and, further, regard them as circular. Thus, we make $\Psi$ cycle around the components of the same triple, instead
of advancing to the next one. Our \rdfcsa then uses a modified array $\Psi$
in which values within $\Psi[2n+1,3n]$ point not to the \subject of the {\em
	next} triple in $T$, but to the \subject of the {\em same}
triple. Thanks to the way we ordered the triples before building $T$, and the grouping of subjects
in $A$, we can compute the modified $\Psi$ very efficiently from the
original array: we simply set $\Psi[i] \leftarrow \Psi[i]-1$ for all positions
corresponding to objects ($i \in [2n+1,3n]$), or $\Psi[i]\leftarrow n$
for the special case $\Psi[i]=1$.

The modified $\Psi$ provides a simpler way to recover and search triples. Since
$\Psi$ cycles over the triples, we can start at any position in the suffix array
$A[i]$, and apply $\Psi$ to recover the remaining components of the triple. For
instance, if $A[i]$ points to a predicate ($i \in [n+1,2n]$), we can find the
object with an iteration using $\Psi$, and the subject with a second iteration
($p = rank_1(D,i)$, $o = rank_1(D,\Psi[i])$, $s = rank_1(D,\Psi[\Psi[i]])$). 
Using the original $\Psi$ we would not be able to
iterate from objects to subjects. Note also that only two iterations are
necessary for any triple, and if we apply $\Psi$ a third time we return to
$i = \Psi[\Psi[\Psi[i]]]$.
The same property allows us to reduce any triple pattern to a search for a short string in $T$. We will further discuss this when describing the query
operations for \rdfcsa.

We note that the modified $\Psi$ used in \rdfcsa, enforcing the property
{$\Psi[\Psi[\Psi[i]]]=i$,} is similar to the {\em permuterm
	index} \cite{Ferragina:2010:CPI}, which tackles a more general case. They
also index a set of strings as if they were circular, so that queries involving
patterns of the form $\alpha*\beta$ (where $*$ stands for an arbitrary string) can be answered by transforming it to the string
pattern $\beta\$\alpha$, where $ \$ $ is a special string
terminator symbol. However, the permuterm index is built on top of an {\em
	FM-index}~\cite{FM05}, which uses a wavelet tree~\cite{FMMN07} as the
underlying data structure.
The wavelet tree implementation requires time logarithmic in the alphabet size, $\mathcal{O}(\log(n_s+n_p+n_o))$ in our case,
for each basic traversal step, equivalent to a computation of $\Psi$ in our
solution. This overhead renders the {\em FM-index} inferior to the \csa on large
alphabets~\cite{FBNCPR12}. We checked this by comparing the best-performing such variant on integer alphabets~\cite{FBNCPR12} to index our sequence $T$, and
obtained times to answer $(s,p,o)$ patterns around $2.5$--$4$ times slower
than those in \rdfcsa. More recent implementations of wavelet trees on large
alphabets have shown only minor improvements for FM-indexes \cite{BCGNNalgor13}. This is
why we implemented our technique on top of the {\em iCSA} for the case of RDF
triples.


\begin{figure*}[htb]
	\begin{center}
		{\includegraphics[width=0.9\linewidth]{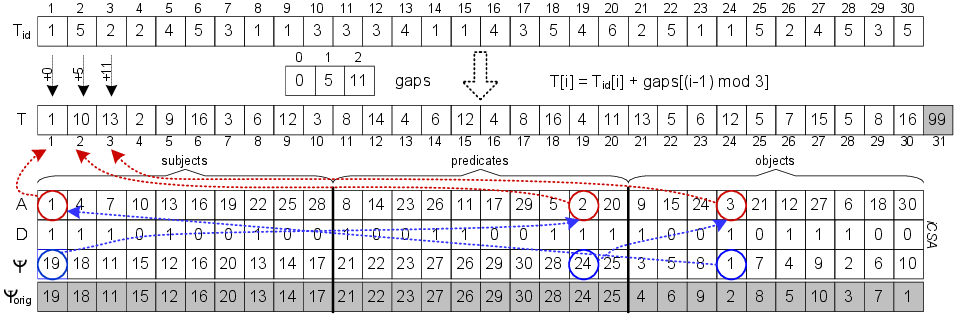}}
		\caption{Structures involved in the creation of a \rdfcsa for the triples in Figures~\ref{fig:rdf} and~\ref{fig:dict}.}
	\end{center}
	\label{fig:rdfcsa}
\end{figure*}

Figure~\ref{fig:rdfcsa} displays the different data structures involved in the
creation of a \rdfcsa for a given set of triples. We use the same triples
described in Figure~\ref{fig:rdf}, following the dictionary encoding of
Figure~\ref{fig:dict}. The collection contains $n=10$ triples, with
$n_s=5$ subjects, $n_p=6$ predicates, and $n_o=5$ objects.
The first step is sorting the triples in $SPO$ order, and concatenating their
components in array $T_{id}$: the first triple is located in $T_{id}[1,3] =
(1,5,2)$, the second one in $T_{id}[4,6]=(2,4,5)$, and so on until the last
triple, which is set in $T_{id}[28,30]=(5,3,5)$.
We compute $gaps[0]=0$, $gaps[1]=n_s=5$, $gaps[2]=n_s+n_p=11$, and then create $T$ by
adding the appropriate component of $gaps[0,2]$ to the values in $T_{id}$. At
the end of this step we obtain $T[1,30]$. Note that we add an extra entry
at the end of $T$ as an implementation trick: by adding this value,
larger than any entry in $T$, we ensure that suffix sorting works properly when
constructing the suffix array $A$, without having to change the construction
used by the original \icsa (similar results could be obtained by adjusting the
algorithm used for suffix comparison).
The suffix array $A$ is then built on top of $T[1,30]$ (recall that the last
element is added to $T$ just for sorting purposes, but it is not considered as
a part of the array itself). Our construction process continues by building the
bitmap $D$ and the array $\Psi_{orig}$ as in the original \icsa. Then, the final array
$\Psi$ used by the \rdfcsa is created from $\Psi_{orig}$ by subtracting $1$ to
$\Psi_{orig}[i]$, for each position $i$ in the interval $[21,30]$ corresponding
to objects, and finally setting $\Psi[30] = 10$ for the special case where $\Psi_{orig}[30] = 1$ (as indicated above).

The main properties stated for $A$ and $\Psi$  be easily checked in the
example. For instance, entries in $\Psi[1,10]$ contain values in the interval
$[11,20]$, entries in $\Psi[11,20]$ contain values within $[21,30]$ and entries in
$\Psi[21,30]$ contain values within $[1,10]$. The figure also displays the
general procedure to traverse the sequence to recover the first triple: starting at
$i=1$, which corresponds to the subject of the triple, we compute
$\Psi[1]=19$ to locate the predicate, and then compute $\Psi[19]=24$ to
locate its object.
Note that if we apply once again $\Psi$, $\Psi[24]=1$ takes us back to the subject location due to the cyclical $\Psi$.
When performing binary search or extracting the
triple, we can get the corresponding values by computing $s=rank_1(D,1)=
\mathbf{1}$, $p = rank_1(D,19)-gaps[1]=10-5 = \mathbf{5}$ and
$o=rank_1(D,24)-gaps[2]=13-11=\mathbf{2}$ to recover the original triple
$(1,5,2)$.

\subsubsection{Data structure optimizations}

The basic implementation described uses the same
data structures as the \icsa~\cite{FBNCPR12} to store $\Psi$ and $D$.
Precisely, $D$ uses the described structures to support $rank$ and $select$, whereas $\Psi$ uses differential encoding combined with Huffman and run-length encoding, which performed best.

On this basic structure, we apply a couple of simple improvements that are specific of the kind of data we are representing. Basically, since the suffix array is separated into three areas of size $n$, for subjects, predicates, and objects, and these have different characteristics, it pays off to separate $D$ and $\Psi$ into three arrays of length $n$ each: $D_s[1,n]$, $D_p[1,n]$, and $D_o[1,n]$, and $\Psi_s[1,n]$, $\Psi_p[1,n]$, and $\Psi_o[1,n]$. We can then encode each array in different form.

In most RDF datasets, the number $|P|$ of different predicates is very small. Since $D_p$ has only $|P|$ 1s, we can avoid the computation of $select_1(D_p,\cdot)$ by directly storing a small array of $|P|$ entries with the results of the $|P|$ distinct $select_1(D_p,\cdot)$ queries;
the $select_1$ operations on $D_s$ and $D_o$ are still carried out as described. The effect in the overall space is negligible.

Further, we add a small structure to speed up $select_1$ queries on $D_s$ and $D_o$: being $n' \le n$ the number of 1s in $D_*$, we add an array ($sOnes$) of $n'/256$ entries where we store the position where every 256th $1$ appears in the bitmap. Given a query $select_1(D_*,i)$, the answer can be either stored in our array (if $i$ is a multiple of 256), or it can be between the samples $\lfloor i/256 \rfloor$ and $\lfloor i/256 \rfloor+1$.  We then start the binary search on the range of the corresponding superblocks, which saves in practice most of the binary search cost. The total space for $rank_1$ and $select_1$ queries is $0.5n$ bits for each of $D_s$ and $D_o$.

The values in $\Psi_s$, which are in $[n+1,2n]$, are decreased by $n$ so that they point inside $\Psi_p$, and those of $\Psi_p$, which are in $[2n+1,3n]$, are decreased by $2n$, so that they point inside $\Psi_o$. These reductions do not affect the differential encodings, but they yield a slight gain of space in the absolute samples, which require $\lceil \log_2 n \rceil$ instead of $\lceil \log_2 3n \rceil$ bits.

More importantly, we can represent each partition of $\Psi$ in different form. We define a variant of our data structure that we call {\em Hybrid}, which slightly increases the space to obtain better access time to $\Psi$. Concretely, {\em Hybrid} stores $\Psi_s$ and $\Psi_o$ in plain form, and keeps
$\Psi_p$ differentially compressed as described. For $\Psi_s$ and $\Psi_o$, we use a simple array requiring $\lceil\log_2 n\rceil$ bits per
entry.
Keeping $\Psi_s$ and $\Psi_o$ uncompressed means that accessing
$\Psi$ will be much faster, in time $\mathcal{O}(1)$ instead of $\mathcal{O}(t_{\Psi})$, in these regions. This will be most noticeable on queries that only use those ranges of $\Psi$.

Choosing a plain representation for $\Psi_s$ and $\Psi_o$ is reasonable because of the characteristics of the \icsa and RDF datasets: the numbers $|S|$ and $|O|$ of different subjects and objects are relatively large, and therefore we take little advantage of the fact that $\Psi_s[1,n]$ and $\Psi_o[1,n]$ are formed by $|S|$ and $|O|$ increasing runs, respectively: this leads to using $\log_2 |S| + \mathcal{O}(\log\log |S|)$ or $\log_2 |O| + \mathcal{O}(\log\log |O|)$ bits to encode each difference, instead of $\log_2 n$ bits to encode an absolute value. For example, using $t_{\Psi}=32$, the differential encoding of $\Psi_s$ reduces its size to 93\% of the plain size using $\lceil \log_2 n \rceil$ bits, and that of $\Psi_o$ reduces it to around 75\%. Instead, because there are few predicates, the differential encoding reduces $\Psi_p$ to around 15\% of its uncompressed size.  This scheme could be easily generalized so as to apply compression only if a given space reduction is achieved.

For simplicity, we will keep speaking of 
$D$ and $\Psi$, ignoring the implementation detail that they are stored in partitioned form.

\subsection{Query operations}
\label{sec:consultas}

In this section we describe how to use \rdfcsa to answer triple-pattern
queries, which constitute the main building block to support SPARQL queries. We describe how to solve the 7 triple-pattern queries $(s,p,o)$,
$(?s,p,o)$, $(s,?p,o)$, $(s,p,?o)$, $(?s,?p,o)$, $(s,?p,?o)$, $(?s,p,?o)$. The basic operatory for 
all of these patterns is to locate the range of entries corresponding to their bound components, and then extracting the corresponding triples.
We will also describe various RDF-specific optimizations.

We disregard the triple pattern $(?s,?p,?o)$, because it retrieves all the triples in the dataset and is not really useful as a query. Nevertheless, we note that it can be easily solved by omitting
the search phase and simply extracting the full set of triples using $\Psi$.

\subsubsection{Solving triple patterns using the regular binary search on
	the \icsa}

The \icsa can locate all the occurrences of a pattern, by binary searching the
range $A[l,r]$ of the suffixes that start with the given pattern. Given a query
pattern $\alpha[1,m]$, the range of positions $[l,r]$ in the suffix array
$A$ will contain pointers to all the positions in the text where the pattern
$\alpha$ occurs. After computing $[l,r]$, $\Psi$ is used to recover the
corresponding symbols.

In our case, we are interested in answering a triple-pattern query, where some
components can be bound and others unbound. As discussed previously, our modified
$\Psi$ allows us to treat all cases similarly, by searching for a subsequence
corresponding to the fixed components in the triple pattern. For instance, to
answer an $(s,p,o)$ query we build a sequence $\alpha[1,3]=spo$, and use that as our 
pattern for the binary search in the \icsa. To answer $(s,p,?o)$ and $(?s,p,o)$
queries, we search for $\alpha[1,2]=sp$ or $\alpha[1,2]=po$, respectively. We can also answer 
$(s,?p,o)$ queries by searching for $\alpha[1,2]=os$, thanks to the cyclical
traversal of our modified $\Psi$. Similarly, for query patterns where only one
of the elements is fixed, we simply search for $\alpha[1,1]=s$, $\alpha[1,1]=p$, or $\alpha[1,1]=o$.
Next we detail the solution for each group of triple patterns, depending on the
number of unbound variables.

For {$(s,p,o)$} queries, we actually set $\alpha[1,3] =
[s+gaps[0],p+gaps[1],o+gaps[2]]$, containing all the elements of the triple
pattern. We then perform a binary search for $\alpha$ in the \icsa. If $l=r$
then $(s,p,o)$ is an existing triple, otherwise it is not in the dataset.

For queries with a single unbound variable, we proceed similarly with a binary
search. Yet, we now have to recover the original triples
afterwards.
For instance, for {$(s,p,?o)$} queries we set $\alpha[1,2] = [s+gaps[0],p+gaps[1]]$. Binary
searching for $\alpha$ in the \icsa, we find the interval $[l,r]$ corresponding
to the result set. The number of answers is $r-l+1$. For each $i \in [l,r]$, we return the triple $(s, p, rank_1(D,\Psi[\Psi[i]])-gaps[2])$.
Similarly, for {$(s,?p,o)$}, we set $\alpha[1,2] = [o+gaps[2],s+gaps[0]]$, then
we binary search for pattern $\alpha$, and return all triples
$(s,rank_1(D,\Psi[\Psi[i]])-gaps[1],o)$. For {$(?s,p,o)$}, we set $\alpha[1,2]
= [p+gaps[1],o+gaps[2]]$, we binary search for $\alpha$, and return the triples
$(rank_1(D,\Psi[\Psi[i]])-gaps[0],p,o)$.

For queries with two unbound variables, we can still perform a binary search to
locate the occurrences of the bound variable. For instance, for {$(?s,p,?o)$}
triple patterns we set $\alpha[1,1]=[p+gaps[1]]$, and find the interval $[l,r]$
with the \icsa. The number of results is again $r-l+1$, and for each $i \in [l,r]$,
the triple $(rank_1(D,\Psi[\Psi[i]])-gaps[0],p,rank_1(D,\Psi[i])-gaps[2])$ is
recovered. Note that, in this case, the binary search in the \icsa does not
require a binary search operation on $\Psi$, since we can compute
$l=select_1(D,\alpha[1])$ and $r=select_1(D,\alpha[1]+1)-1$. As in the previous
examples, {$(?s,?p,o)$} and {$(s,?p,?o)$} can be answered using exactly the same
operation but adjusting $\alpha$ and the computation to return the result
triples.

%
%
%
	%
	%

Since we are using a binary search on the \icsa, all the triple-pattern queries
require $\mathcal{O}(r-l+\log n)$ time, where $r-l+1$ is the number of query results.
In addition to this, for most query patterns we need to perform a
number of accesses to $\Psi$ per query result in order to return the
complete triples. 
In practice, efficient access to $\Psi$ must be balanced
with efficient compression; the compression of $\Psi$ introduces a significant
space/time tradeoff that can be tuned in our representation. Note that
the space/time tradeoff also depends on the type of query pattern involved: if
a query returns a large number of results, the cost of the binary search becomes
negligible and the time required to perform accesses to $\Psi$ dominates the
cost of the query. However, the binary search cost becomes relevant when only
one or a few triples are returned, as well as in $(s,p,o)$
queries, where no triple-pattern retrieval is necessary.

\subsubsection{Query optimizations} 
\label{sec:variantes}

We now describe a number of optimizations and algorithmic variants that 
improve our performance.

One enhancement improves query patterns with two unbound
terms, in which we always need to perform two $select$ operations on $D$ over
two consecutive values, $i$ and $i+1$.
Once we compute $j = select_1(D,i)$, we can replace $select_1(D,i+1)$ by a new operation $selectnext(D,j)$, which finds the next 1 after $D[j]$. We implement $selectnext$ by scanning $D$ bytewise from position $j+1$ to the end of its block. If we find no 1 up to then, we scan the following 32-bit words looking for a nonzero block. If we find no 1 up to then, we check if the next superblock has a $1$, and if not, we binary search for the next one that has. On that superblock, which contains the answer, we restart the wordwise scan, then the bytewise scan, and finally use the same table of $select$ to find the desired $1$.
This is in practice faster than a second binary search.
%
%

Our next optimization improves the performance of accesses to
$\Psi$, particularly taking into account that in most cases we need to compute
values of $\Psi$ for a relatively large range of consecutive positions.
In the original algorithm, once $[l,r]$ is determined through binary search, we 
have to compute $\Psi[i]$ and $\Psi[\Psi[i]]$ for all $i \in [l,r]$ to
retrieve the missing elements in each triple (except on the pattern $(s,p,o)$).
Since $\Psi$ is differentially encoded, each access takes time $\mathcal{O}(t_\Psi)$, where we spend $(n\log n)/t_\Psi$ bits to store the absolute samples.
In order to improve the speed of these accesses, we
sequentially decompress the whole range $\Psi[l,r]$. This means that, once we decode $\Psi[l]$ in $\mathcal{O}(t_\Psi)$ time,
all the subsequent values are decoded in constant time. This variant is particularly efficient if we are inside a run of differences equal to $1$, as these are encoded using run-length encoding. Note that this only works for the initial
range $[l,r]$, since the remaining accesses to $\Psi$ are expected to be
located at random and therefore they cannot be improved with this technique.
	%
	%

We also improve the strategy to binary search for $[l,r]$. We describe two alternative strategies,
called {\em D-select+forward-check} and {\em D-select+backward-check}, which apply to patterns with 2 or 3 bound elements.

\paragraph{\bf{D-select+forward-check strategy}} During a binary search in the
\icsa, we compare the query pattern $\alpha$ with the string pointed by the
current position in the suffix array, $T[A[i],n]$. The first steps of the binary search will be faster because the strings will differ in their first
character, so the comparison will be decided with the first integer comparison without the need to compute $\Psi$, just  
$T[A[i]]=rank_1(D,i)$. At some step
of the binary search, however, we will start to have $T[A[i]]=\alpha[1]$ and will have to compute
$\Psi[i]$ in order to compare $\alpha[2]$ with $rank_1(D,\Psi[i])$; this access
to $\Psi$ can be relatively expensive if differentially compressed.

Instead of performing all those isolated $\Psi$ computations, in
this strategy we perform all the
checks for the complete range in order to filter the candidate
positions.

Consider for instance the triple pattern $(s,p,o)$, in which we would search for $\alpha=spo$.
We first find the intervals that correspond to the subject,
predicate, and object of the triple pattern: $R_s=[l_{s+gaps[0]},
r_{s+gaps[0]}]$, $R_p=[l_{p+gaps[1]}, r_{p+gaps[1]}]$, and $R_o=[l_{o+gaps[2]}, r_{o+gaps[2]}]$, using $select$ operations on $D$: $l_c =
select_1(D,c)$ and $r_c = selectnext(D,l_c)-1$. Since $\Psi$ is increasing within
each of those intervals, we use these ranges to check, for each $i$ in $R_s$, whether $\Psi[i] \in R_p$. Only a smaller range $R_{sp} \subseteq R_s$ will pass this filter, and the $\Psi$ values in that range form in turn a range $R_{ps} \subseteq R_p$. On this range $R_{ps}$ we compute all the $\Psi$ values to finally find the range $R_{pso} \subseteq R_{ps}$ of the values that map inside $R_o$ by $\Psi$. Those are the final answer.

\begin{figure}[htb]  
	\begin{center}
		{\includegraphics[width=1.0\linewidth]{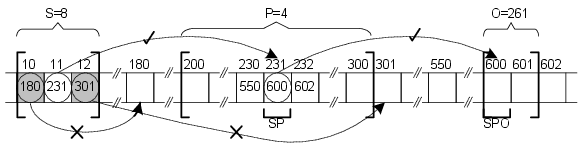}}
		\caption{D-select+forward-check strategy for pattern $(s,p,o)=(8,4,261)$.}
		\label{fig:selectDForward}
	\end{center}
\end{figure}

Figure~\ref{fig:selectDForward} shows an example of this operation. In this
example, $R_s=[10,12]$, $R_p=[200,300]$, and
$R_o=[600,601]$. Checking the values of $\Psi$ for the range $R_s$, we find that 
$\Psi[10]$ and $\Psi[12]$ do not map into range $[200,300]$, but $\Psi[11]$
does. Therefore, we need to check if $\Psi[\Psi[11]]$ maps into the
range $R_o=[600,601]$, corresponding to object $261$. Since it matches,
we can report an occurrence of the triple $(8,4,261)$, i.e., confirm that the
triple is in the collection.

In practice, this technique may be faster than a standard binary search
if the initial interval ($R_s$ in our example) is small enough.
Note that, since our $\Psi$ is cyclic, we can use any of the three
intervals $R_s$, $R_p$, or $R_o$ to begin our check.
Typically, the number of objects is higher than that of subjects, so we expect that $\lvert R_o\rvert < \lvert R_s\rvert \ll \lvert R_p\rvert$. We may, however, choose on the fly the one that is actually shortest.

The strategy presented here can also be applied to triple patterns with one
unbound term. In this case, we perform the same operations but restricted to the
bound terms. Assuming our bound variables are $x$ and $y$, we compute $R_x$ and
$R_y$ and perform the same range check to verify if, when applying $\Psi$ to the
positions in $R_x$, we end up in range $R_y$. Again, notice that the cyclic
nature of $\Psi$ allows us to perform the range check independently of the position of
the bound variables in the triple pattern. For example, for $(?s,p,o)$ triple patterns
we set $x=p$, $y=o$; for pattern $(s,?p,o)$, we set $x=o$, $y=s$; and for
pattern $(s,p,?o)$ we set $x=s$, $y=p$.


\paragraph{\bf{D-select+backward-check strategy}} This strategy is based on the
same ideas of the previous forward-check strategy. It relies on the fact that all positions $i$ in $R_s$ that pass the
forward-check in the previous strategy necessarily form a subinterval of $R_s$.
This means that, in order to discard candidate positions, we do not need to
verify every $i \in R_s$; instead, we can binary search for the subrange of
positions that map to a valid range in $R_p$.

To take advantage of the previous property, we follow a similar idea to the
well-known backward-search strategy~\cite{Sad03}.
Assume that we are searching for a triple pattern $(s,p,o)$. We start our search
now in interval $R_o=[l_o,r_o]$; since $\Psi$ must be increasing within interval
$R_p=[l_p,r_p]$, we binary search inside $R_p$ in order to locate the
subinterval $R_{po}=[l_{po}, r_{po}]\subseteq R_p$ that contains all the
positions $i$ such that $\Psi[i] \in R_o$. If the subinterval is empty, no
result exists for the query and we return immediately. Otherwise, we continue
the backward-search process, binary searching in $R_s$ in order to locate the
subinterval $R_{spo} = [l_{spo},r_{spo}] \subseteq R_s$ that contains all
the entries $i \in R_s$ such that $\Psi[i] \in R_{po}$. At the end of this step,
the range $R_{spo}$ contains all the results for our query. Note that, when
using an $(s,p,o)$ pattern, either 0 or 1 results may arise, but we generalize this
strategy to other triple patterns below.

\begin{figure}[htb]  
	\begin{center}
		{\includegraphics[width=1.0\linewidth]{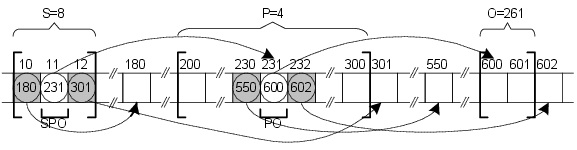}}
		
		\caption{D-select+backward-check strategy for pattern $(s,p,o)=(8,4,261)$.}
		\label{fig:selectDBack}
	\end{center}
\end{figure}

Figure~\ref{fig:selectDBack} displays an example of this strategy for a
sample $(s,p,o)$ query pattern. We start the backward search in range $R_o =
[600,601]$. Then we perform a binary search in the interval $\Psi[200,300]$, in
order to locate the subinterval that contains values that map into $R_o$; in our
example, only the entry $\Psi[231]$ maps into $[600,601]$, so we obtain a
subinterval $R_{po} = [231,231]$. Next, we continue the backward-search in
$R_s$. We binary search inside the range $\Psi[10,12]$ and locate the
subinterval that maps to $231$; in the example, only $\Psi[11] = 231$ maps. Consequently, the
final interval is $R_{spo}=[11,11]$, that contains the single occurrence for
the given pattern.

This strategy can be easily adapted to work with all the
query patterns that contain a single unbound variable.
In $(s,p,?o)$ queries, we locate the subinterval $R_{sp} \subseteq
R_s$ that maps into $R_p$ after applying $\Psi$. In $(s,?p,o)$ queries, we
locate the subinterval $R_{os} \subseteq R_o$ whose $\Psi$ entries map
into $R_s$. In $(?s,p,o)$ queries, we locate the subinterval
$R_{po} \subseteq R_p$ whose entries map into $R_o$.

\subsection{Supporting join operations}\label{sec:joins}

\rdfcsa can be extended to support join operations by implementing different join techniques on top of the basic triple pattern query algorithms. We first describe the general technique, which can be
used with any number of unbound elements in the triple patterns and for
subject-subject, subject-object, and object-object join operations. We then 
briefly explain particular optimizations that are applied on top of the
general technique.

Join operations in \rdfcsa are essentially performed by following either a {\em
	merge-join} strategy or a {\em chaining} strategy.

The merge-join strategy considers each triple pattern separately. The
join variable is treated as an unbound variable in both triple patterns. The two
corresponding triple patterns are solved independently, therefore obtaining two
lists of results.
The final step scans the resulting lists to compute their
intersection.\footnote{Since the results returned by the \rdfcsa for some triple
	patterns are not necessarily sorted by the desired element, a sorting step may be required prior to the intersection.}
For instance, to compute $(s_1,p_1,?x) \bowtie (s_2, p_2, ?x)$, we first compute the two triple-pattern queries $Q_1 = (s_1,p_1,?o_1)$ and $Q_2 = (s_2, p_2, ?o_2)$. The results of
$Q_1$ and $Q_2$ are then intersected by the $O$ component to retrieve only the values where $o_1=o_2$. The same strategy can be applied to any
combination of triple patterns, with simple adjustments depending on the number of unbound variables
in each side.

The chaining strategy, instead, solves one of the triple patterns first, considering the
join variable as unbound. Then, for each result obtained in this
query, the second pattern is executed with the corresponding value of
the join variable, which is now bound. The previous example, $(s_1,p_1,?x) \bowtie (s_2,
p_2, ?x)$, is executed following this strategy by first querying
$(s_1,p_1,?o_1)$, and then replacing each value $o_1$ obtained for $?o_1$ in the
second pattern as $(s_2, p_2, o_1)$. We speak of left-chaining if we start with the left triple pattern and apply each result as bound variables in the right one (as in the previous example), and of  right-chaining if we start executing the right triple pattern and replace the results in the left one. The selection of the first pattern for chaining is important when the triple patterns have a different number of unbound variables.

In \rdfcsa we have implemented a general mechanism to perform joins following the merge
strategy as well as a left- or right-chaining strategy. 
Depending on the
characteristics of the join, and particularly the location of the unbound
variables, the strategy selected leads to significantly different triple-pattern queries, and therefore to important differences in query performance.
The selection of the optimal strategy is therefore a significant problem by
itself. We test all possible strategies in our experimental evaluation,
with one exception: strategies that would lead to the evaluation of an
$(?s,?p,?o)$ pattern as a first step are not considered in any case, since
decompressing the full dataset as an intermediate result would be very
inefficient in terms of time and space.

\subsubsection{Optimization of join operations}

Some optimizations are added on top of the general join strategy, to take
advantage of the characteristics of our technique and specific join patterns. These optimizations have a significant effect on the amount of computation performed by \rdfcsa in most join operations.

The first enhancement to the basic algorithms is related to the
dictionary encoding used. Recall that in the dictionary encoding used by HDT,
all elements that are both subject and object are assigned an id lower than
that of any element that only appears as a subject or as an object. This can be used
to filter out results when performing subject-object joins. For example, to
answer a query $(s_1,p_1,?x) \bowtie (?x,p_2,o_2)$ using left-chaining, we would
first obtain all the objects that match the triple pattern $(s_1,p_1,?x)$;
then, we have to check that each result matches the right triple pattern.
However, with the dictionary encoding we use, we can immediately discard any
result of the first query with an id higher than $|SO|$, since we know that it
only appears as an object and therefore it will not match the overall join
query. Note that this improvement is specific to this dictionary encoding, and
is not specific to \rdfcsa; the same optimization is also used, for instance,
in \kdostriples.

Another simple optimization that is applied to the merge strategy
consists in taking into account the characteristics of the result list returned.
In some join patterns, we must sort both lists to compute their
intersection; however, due to the evaluation mechanisms of \rdfcsa, in some
triple patterns the list of results is already sorted. For instance, the
$(s,p,?o)$ triple pattern returns a sorted list of objects as a result;
therefore, to answer a query $(s_1,p_1,?x) \bowtie (s_2,p_2,?x)$, we can execute
the two triple-pattern queries and then simply intersect the corresponding
sorted lists. A similar idea is also applied to the chaining strategy: we can avoid some
computation in the chaining phase by identifying repeated results. In order to
do this, we sort the results of the first triple pattern and skip the
computation of the second triple pattern on the repeated results of the first
query. Therefore, we build the results of the final join only from the
non-repeated results of the first triple pattern.

An additional improvement we include in all our join operations, when possible,
is \emph{variable filling}. As explained before, when running most triple-pattern
queries, we first obtain the location of the set of triples and then use
$\Psi$ to retrieve the missing variables in the triple. This cost is
necessary to return the complete result in a triple-pattern query. However, in join queries that follow the merge or
chaining strategy, many of the matches found in the first pattern may not
correspond to valid results of the overall join operation, since they do not
have a match for the join variable in the second pattern.
Our algorithms identify, depending on the type of join and the
evaluation strategy, which variables in a triple pattern are necessary to
solve the join and which ones are only necessary to make up the final result.
The latter variables are filled in only after the
complete join has been evaluated.  We then use slightly modified versions of each
triple-pattern query, customized according to which of the
elements in the resulting triple have to be computed. The general algorithms
solve the join using the incomplete
triples (hence avoiding the rather costly $\Psi$ computations on non-sampled
positions, and rank operations), and then take care of refilling the missing
variables once the join has been completed.

For instance, to perform the
join $(?s_1,p_1,?x) \bowtie (?x, p_2, o_2)$ with left-chaining, the first step is to compute the left triple pattern $(?s_1,p_1,?o_1)$. This is usually done by
first locating the range of $p_1$, and then using $\Psi$ to
locate the corresponding objects, and $\Psi$ again to get the subjects. However,
for the join operation we do not need the subjects, only the objects, so we do not compute
the subjects yet: we first complete the join query, and then fill in
the missing subjects for the resulting tuples.

%


\section{Experimental evaluation} 
\label{sec:experiments}

\subsection{Experimental framework}


We tested the compression and query performance of our proposal using
the {\em DBPedia} dataset,\footnote{\url{http://downloads.dbpedia.org/3.5.1/}}
``the nucleus for a Web of Data''~\cite{ABKLCI:07}. The original size of the
dataset is around 34GB. It contains 232,542,405 triples in total, 18,425,128
different \subjects, 39,672 different \predicates, and 65,200,769 different \objects.
After applying dictionary encoding to the triples, the structural
part of the dataset can be stored in 2,790,508,860 bytes, using three 32-bit
integers per triple.

We compare \rdfcsa with \kdostriples and permuted trie indexes, as good examples of other well-known state-of-the-art solutions that are similar to \rdfcsa, in the sense that they are based on compact data structures and designed to work with triples composed of integer identifiers. We also compare our proposal with a number of alternative solutions following other approaches: HDT, Tentris, Virtuoso (open source edition, version 7.2.5.1), Blazegraph (version 2.1.4), MonetDB (version 1.7), and RDF-3X (version 0.3.7). Note that all of the latter can handle the RDF datasets in their original form as string triples. In the case of HDT, we display in the plots the space required only for the Triples component, so it is directly comparable to \rdfcsa, \kdostriples, and permuted trie indexes. The same occurs for MonetDB, where we also store and query integer identifiers instead of strings. For Tentris, Virtuoso, Blazegraph, and RDF-3X we display the full size of the structure, after loading the original RDF dataset. The effect of the dictionary in space and query times will be discussed later. 

Regarding query times, measurements are also taken differently in each family of solutions. For Virtuoso, Blazegraph, Tentris, MonetDB, and RDF-3X we measure query times using the utilities provided by each tool, which includes, in general, the cost of parsing the query. For \rdfcsa, \kdostriples, permuted trie indexes, and HDT, we measure the performance of queries on the integer ids, therefore ignoring any additional costs associated to the query tool and the SPARQL query parsing necessary in the other solutions. Query times are always displayed in $\mu$s/result to reduce the effect of the overhead required by the more complex tools.

For \rdfcsa, we test the different algorithms and variants using different sampling intervals on $\Psi$, 
$t_{\Psi} \in \{4,8,16,32,64,512\}$, so as to obtain a wide
space/time tradeoff. Additional details on the variants and configurations will be given later.

For \kdostriples we use the settings recommended by the authors. 
We use two different implementations, the original \kdostriples and the improved \kdostriplesplus that includes extra indexes to speed up queries with unbound predicate. 

We test two configurations of the permuted trie index:\footnote{\url{https://github.com/jermp/rdf_indexes}} \trieb and \trietwo. The former has
better performance and offers more stable query times because it is efficient over all triple patterns. Instead,
\trietwo uses only two of the three permutations, so as to reduce space
while maintaining query times in most 
triple patterns. The
main drawback of \trietwo is that it performs much worse on
$(?s,?p,o)$ triple patterns. There are other configurations of the permuted trie index, but we have chosen the best performing ones according to its authors.

For HDT, we use the original implementation by the authors.\footnote{\url{http://www.rdfhdt.org/}} To provide
comparable query times, we performed minimal changes to the source code in
order to measure only the structural part of the query. To do this, we
precompute the string-to-id translation for all queries, and then measure query
times to return all results as identifiers, omitting the final id-to-string
translation that is usually performed to return the final results. Therefore, our plots reflect the space and time required to solve the query on ids, omitting the space and time required for the HDT dictionary.

For MonetDB,\footnote{\url{https://www.monetdb.org/}} we store the integer ids corresponding to the triples to make their results directly comparable to the previous solutions. We use the mclient command-line tool to execute queries, and use the query times reported by the tool.

For Virtuoso,\footnote{\url{https://virtuoso.openlinksw.com/}} we use the ingestion and query tools provided with the software. Particularly, we use the interactive command-line query tool isql to execute queries, and use the query times reported by the tool. Note that Virtuoso includes a server that provides an HTTP endpoint that can be used to run SPARQL queries. We have also tested query times in this interface, but the overhead caused by this endpoint was very significant (query times were 1.3--3 times larger than in the command-line tool); additionally, the HTTP endpoint limits the number of results returned, making it impractical for our purposes.

For Blazegraph,\footnote{\url{https://blazegraph.com/}} we use a custom Java program that connects to Blazegraph in embedded mode. Query times are measured using \textsf{System.nanoTime}.
The SPARQL endpoint provided by Blazegraph was also tested, but the overhead caused by it was also significant.

For Tentris, we use the query tool tentris\_terminal,\footnote{\url{https://github.com/dice-group/tentris}} provided by the authors. We display query times as measured by the tool. Since parsing times are disgregated by the tool, we display two different times for Tentris: \textsf{Tentris} represent total times, whereas \textsf{Tentris-noparse} exclude the parsing time from the total. Again, we have tested the HTTP endpoint provided, but we omit these results in our plots because they were up to 10 times worse than the command-line results in some queries.

In RDF-3X,\footnote{\url{https://code.google.com/archive/p/rdf3x}}  we use the command-line query tool provided to run the queries and measure query times.

We use an existing testbed for the DBPedia dataset.\footnote{Provided by the authors of 
	\kdostriples, available at
	\url{http://dataweb.infor.uva.es/queries-k2triples.tgz}} This query set provides 500 queries for each of the 7 basic triple patterns, and 25 queries for each join pattern considered (additional details on the join variants and their classification will be provided in Section~\ref{sec:exp:joins}). For a fair comparison with tools that require access to disk, we execute a warm-up phase before running each query set. The warm-up includes performing the full set of triple pattern queries. After that, we execute each query set, measuring query times. Additionally, we set a number of repetitions of the full query set for triple pattern queries to guarantee accurate
average time measurements.

We ran our experiments on an Intel Xeon E5-2470@2.3GHz (8 cores)
CPU, with 64GB of RAM. The operating system was Debian 9.8 (kernel
4.9.0-8-amd64). The version of GCC was 6.3.0, and the version of Java (used to run Blazegraph) was 1.8. Our code, as well as the source code for RDF-3X and HDT, were compiled using GCC, with full optimizations. The remaining tools were installed using the packages/binaries provided by the authors. 

We have made our source code available at \url{https://lbd.udc.es/research/rdf/}.

\subsection{Comparison of the query algorithms of \rdfcsa}

First we analyze the relative performance of the query algorithms developed for
our structure, discussed in Section~\ref{sec:consultas}. We measure space and
query times for the different triple patterns using the basic binary search algorithm (\emph{base} in the plots), the
D-select-forward-check strategy (\emph{forward}), and
D-select-backward-check (\emph{backward}). 

\begin{figure*}[t!]
	\centering
	\includegraphics[angle=-90,width=0.3245\textwidth]{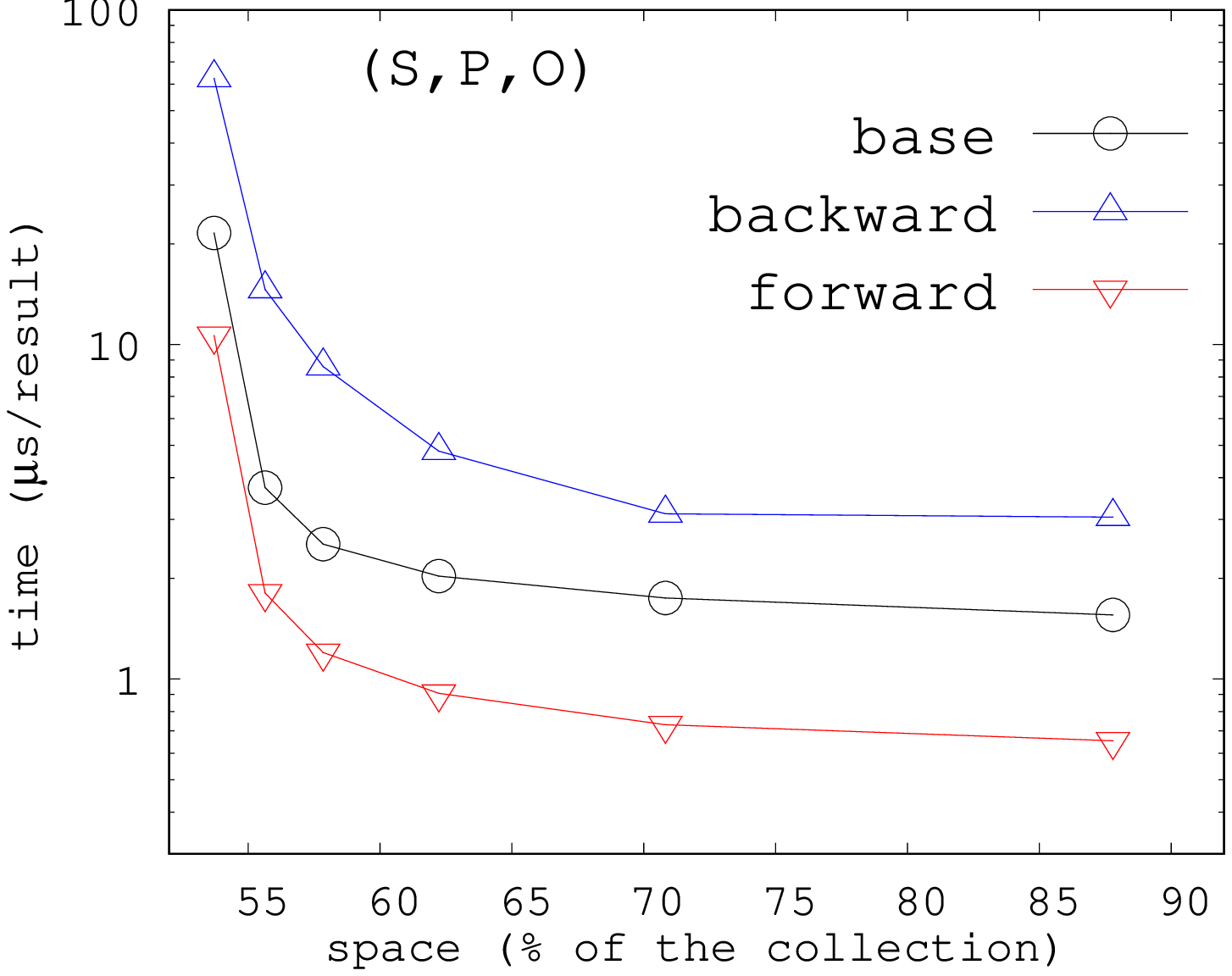}
	\includegraphics[angle=-90,width=0.3245\textwidth]{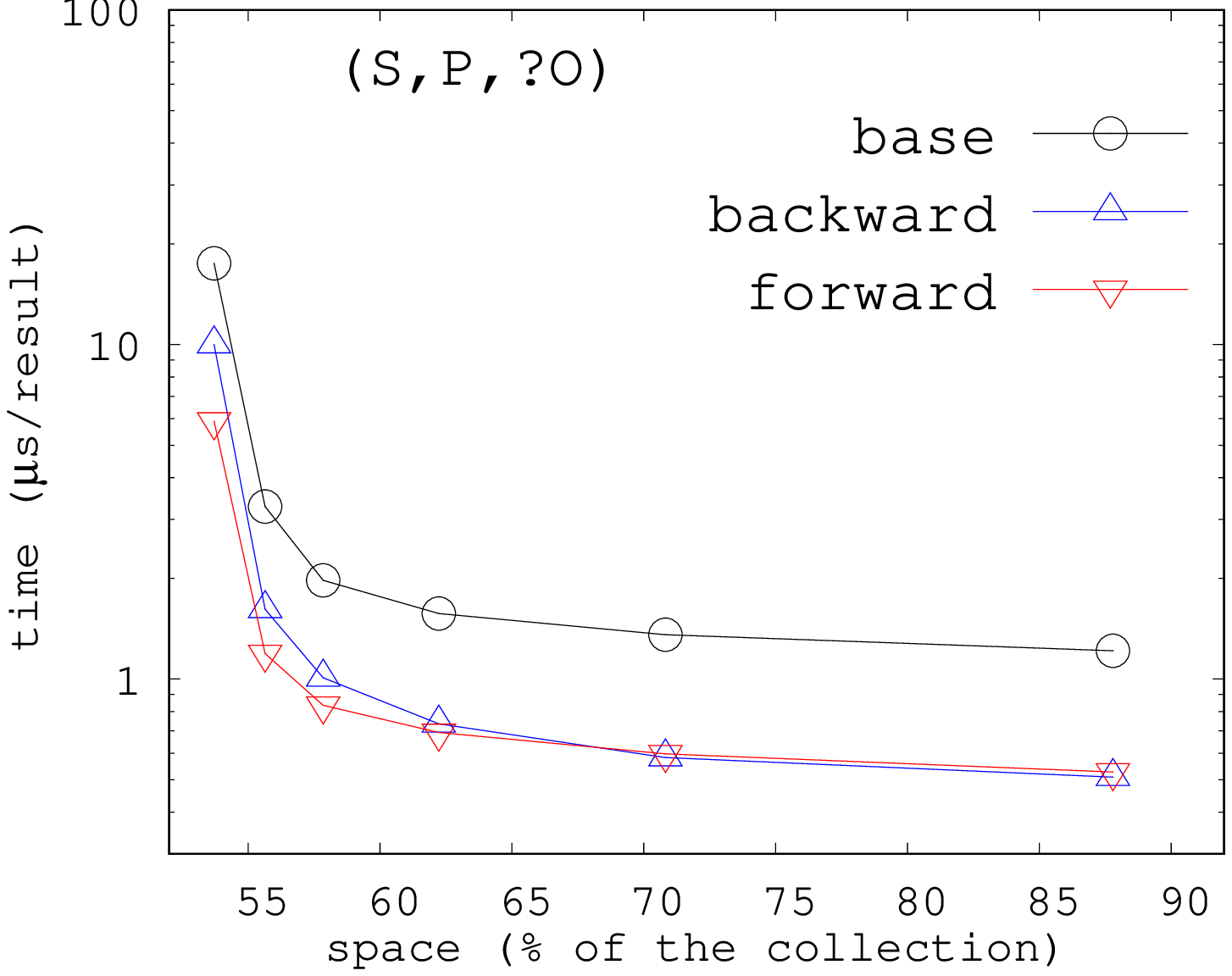} \\
	\includegraphics[angle=-90,width=0.3245\textwidth]{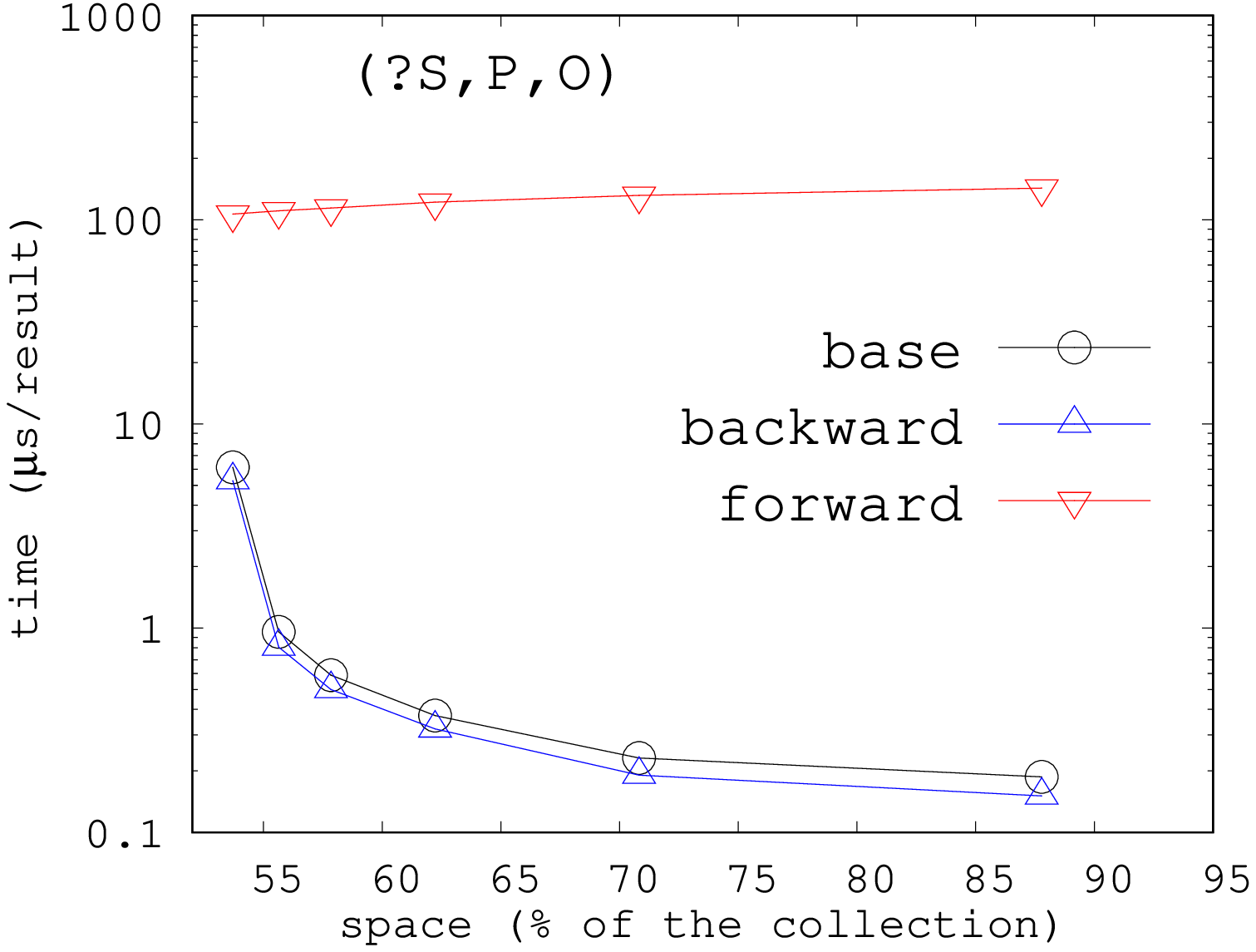}
	\includegraphics[angle=-90,width=0.3245\textwidth]{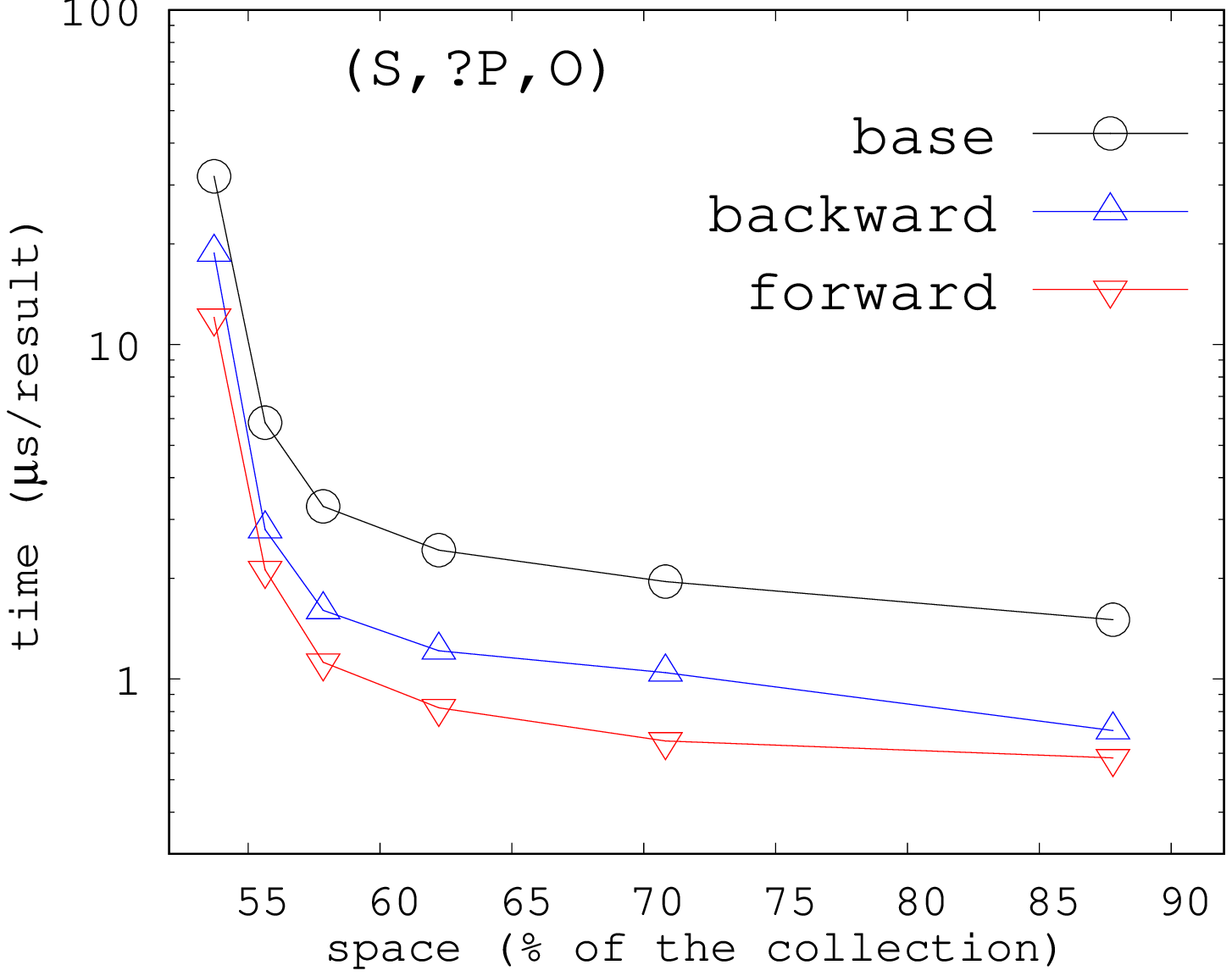}
	\caption{Query times of the search variants on query patterns with
		zero or one unbound variable. Times in microseconds per result returned and in log scale.}
	\label{fig:iexperiments}
\end{figure*}

\begin{figure*}[t]
	\centering
	\includegraphics[angle=-90,width=0.3245\textwidth]{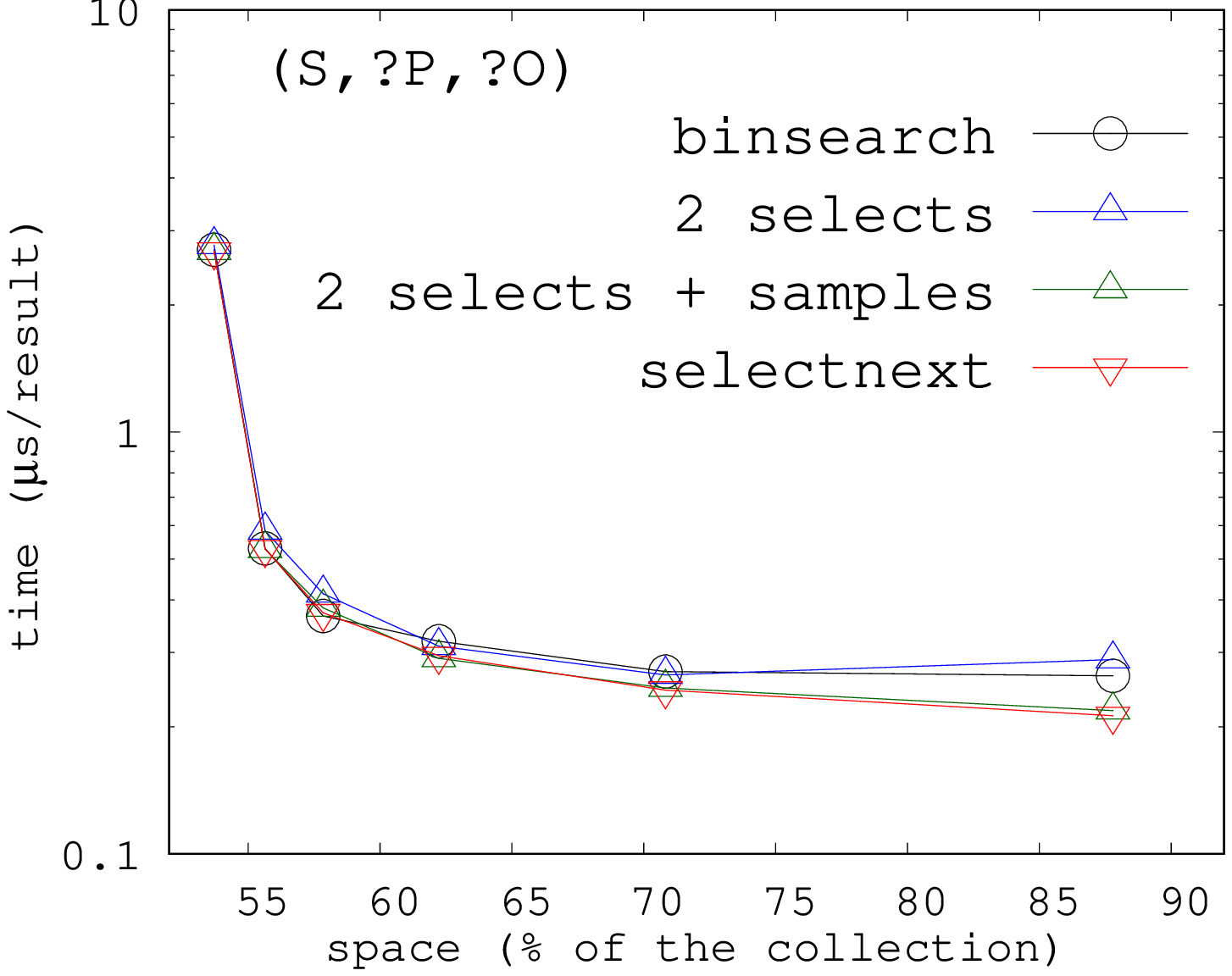}
	\includegraphics[angle=-90,width=0.3245\textwidth]{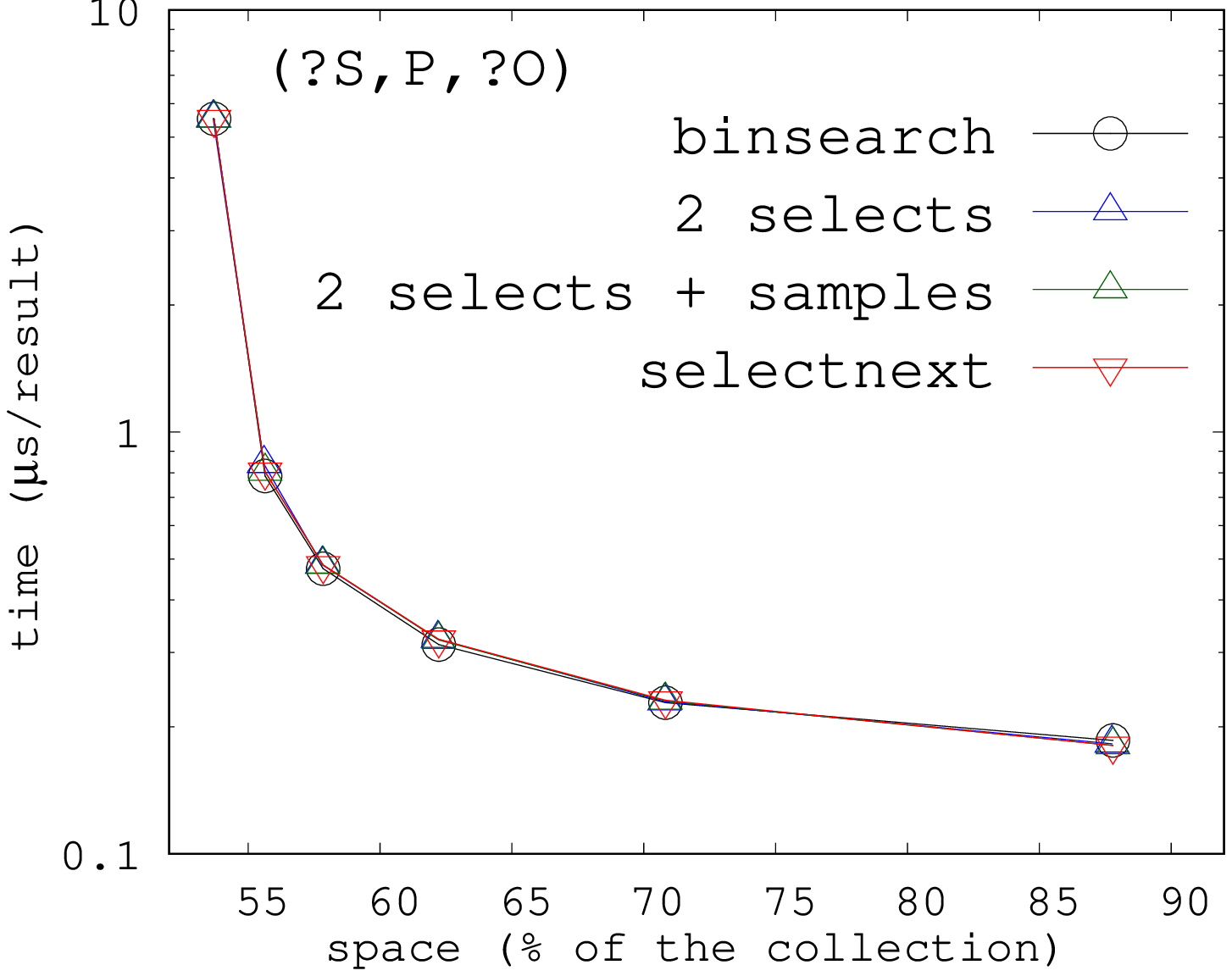}
	\includegraphics[angle=-90,width=0.3245\textwidth]{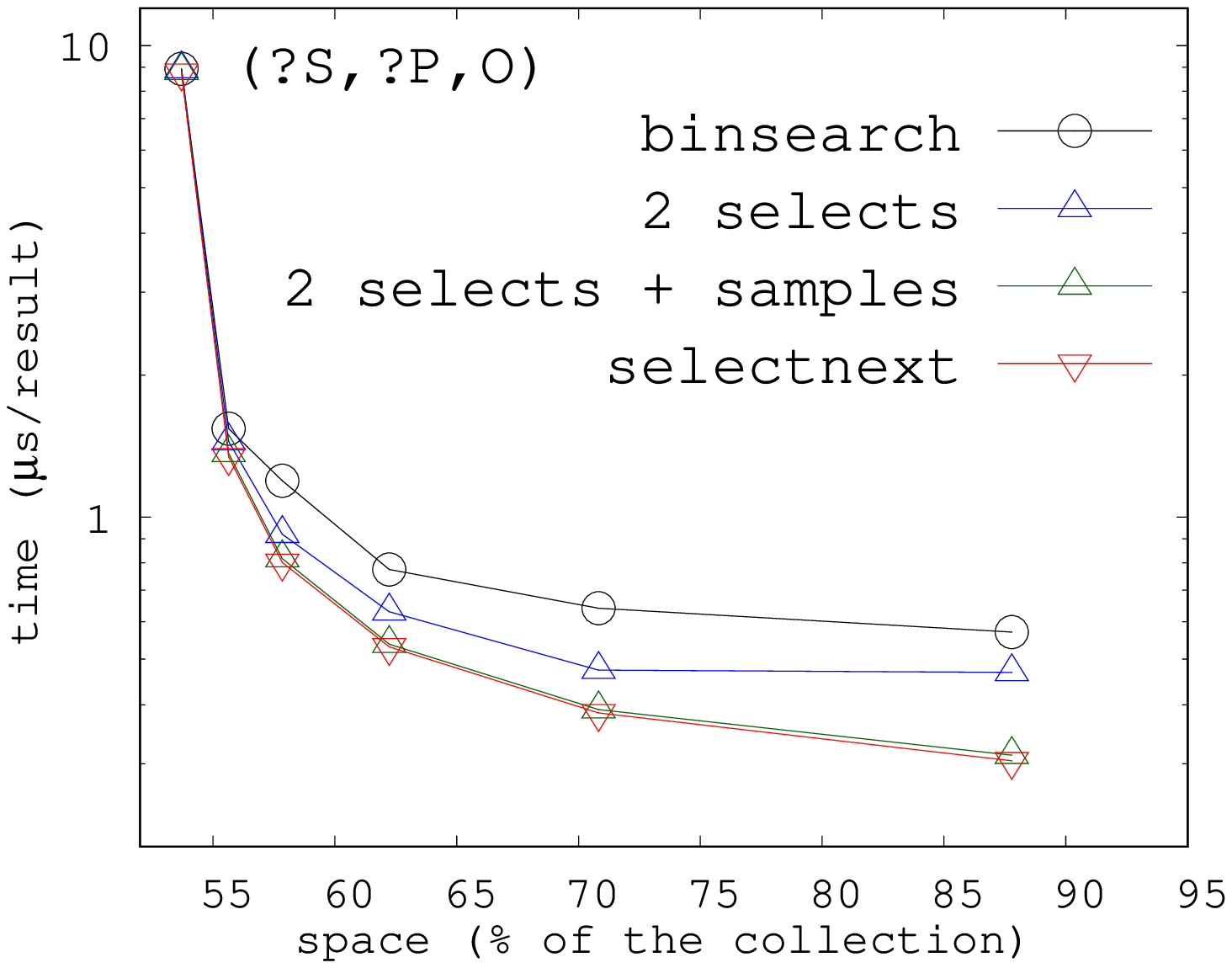}
	\caption{Comparison between basic binary search in the \icsa and dual select
		for patterns with one fixed term. Times in microseconds per result and in log scale.}
	\label{fig:biselect}
\end{figure*}

Figure~\ref{fig:iexperiments} displays the space and query times for the
different search algorithms.\footnote{The space is given as a percentage of the size of the raw data, which for this purpose is taken as a binary
	representation of the triple patterns with  each triple stored using
	three 32-bit integers.} We only show results for query
patterns with zero or one unbound variable, because triple patterns with a single fixed variable lead to patterns $\alpha$ of length 1, where backward- or forward-check
strategies cannot be applied. For the backward- and forward- strategies we use our $selectnext$ optimization.\footnote{Further details comparing {\em select} implementations will be given in Figure~\ref{fig:biselect}.}
As shown in the figure, the baseline binary search is in general slower 
than the other alternatives. 
A notable exception occurs in $(?s,p,o)$ queries, where the forward-check strategy is very inefficient. This difference is due to the large number of occurrences that may have to be sequentially checked in $R_p$. Therefore, even though {\em
	D-select+forward-check} is faster in most cases, {\em D-select+backward-check}
is in general more consistent. Note, nevertheless, that we can easily select
the best algorithm for each triple pattern, and we can even perform on-the-fly
selection of the best query algorithm using a simple heuristic depending on the
length of the ranges involved. For simplicity, in the following
experiments we only display the query time of the most efficient search
technique in each query pattern (i.e., {\em D-select+forward-check} in most cases,
{\em D-select+backward-check} in $(?s,p,o)$ queries). Note also that the
results presented in this section are those of the basic implementation of \rdfcsa. Additional
plots are omitted for simplicity, but we have obtained similar results for other
implementation variants, with {\em D-select+backward-check} being the most
consistent search strategy overall.

\medskip
Next, we analyze the impact of our improvements on $select_1$ queries on triple patterns with 
two unbound variables. In these queries, we must search for a
pattern $\alpha$ of length 1, so we can replace the
standard binary search of the \icsa by two $select$
operations in $\Psi$ to locate the appropriate interval $[l,r]$. Further, the second $select$ can be replaced with the $selectnext$ algorithm, which is faster (see Section~\ref{sec:variantes}).

Figure~\ref{fig:biselect} displays the performance of the binary search on $\Psi$ ({\em binsearch}), of replacing it with two $select_1$ operations on $D$ implemented with binary searches ({\em 2 selects}), of improving those $select_1$ operations with sampling ({\em 2 selects $+$ samples}), and of replacing the second such $select_1$ with a $selectnext$ operation ({\em selectnext}). The results show that each improvement makes a significant difference with the previous version, except for the use of $selectnext$, whose improvement is marginal but still always positive. Recall that we store the $select_1$ answers directly on $D_p$, thus in the triple pattern $(?s,p,?o)$ there is no difference between {\em binsearch} and the various $select_1$ variants. 
Considering these results, in the remaining experiments we will always use the
$selectnext$ algorithm when applicable.

\subsection{Comparison with other RDF representations}\label{subsec:exp-other}

In this section we compare \rdfcsa with state-of-the-art alternatives. We start
by measuring their space requirements and query performance on simple triple patterns.
We show compression as a percentage of the original size of the
collection (considering an integer-base representation). We test three implementation variants
of \rdfcsa. In all of them, we use the algorithms that obtained the best results in previous tests:
$selectnext$ to obtain ranges using $D$,  {\em
	D-select+forward-check} for most patterns that require search on $\Psi$, and {\em D-select+backward-check} for
$(?s,p,o)$ patterns. The three variants of RDFCSA tested are the following:

\begin{itemize}
	\item \emph{RDFCSA} is the basic implementation, with $D$ and $\Psi$ partitioned into three arrays. Those for $D$ are bitmaps in plain form with $rank_1$ \cite{GGMN05} and our faster $select_1$ structures, yet $D_p$ stores the $select_1$ answers in plain form. The $\Psi$ arrays are  compressed with Huffman and run-length encoding (RLE)~\cite{FBNCPR12}.
	\item \emph{RDFCSA-rrr} is like the basic variant but the bitmaps of $D$ are compressed using the RRR
	technique~\cite{RRR07} with sampling parameter 128.
	\item  \emph{RDFCSA-Hybrid} is the \emph{hybrid} variant, with $\Psi_s$ and
	$\Psi_o$ stored as plain arrays where entries use $\lceil \log_2 n \rceil$
	bits, and $\Psi_p$ compressed as usual with Huffman and RLE.
\end{itemize}

\begin{figure*}[!t]
	\begin{center}
		\includegraphics[angle=-90,width=0.33\textwidth]{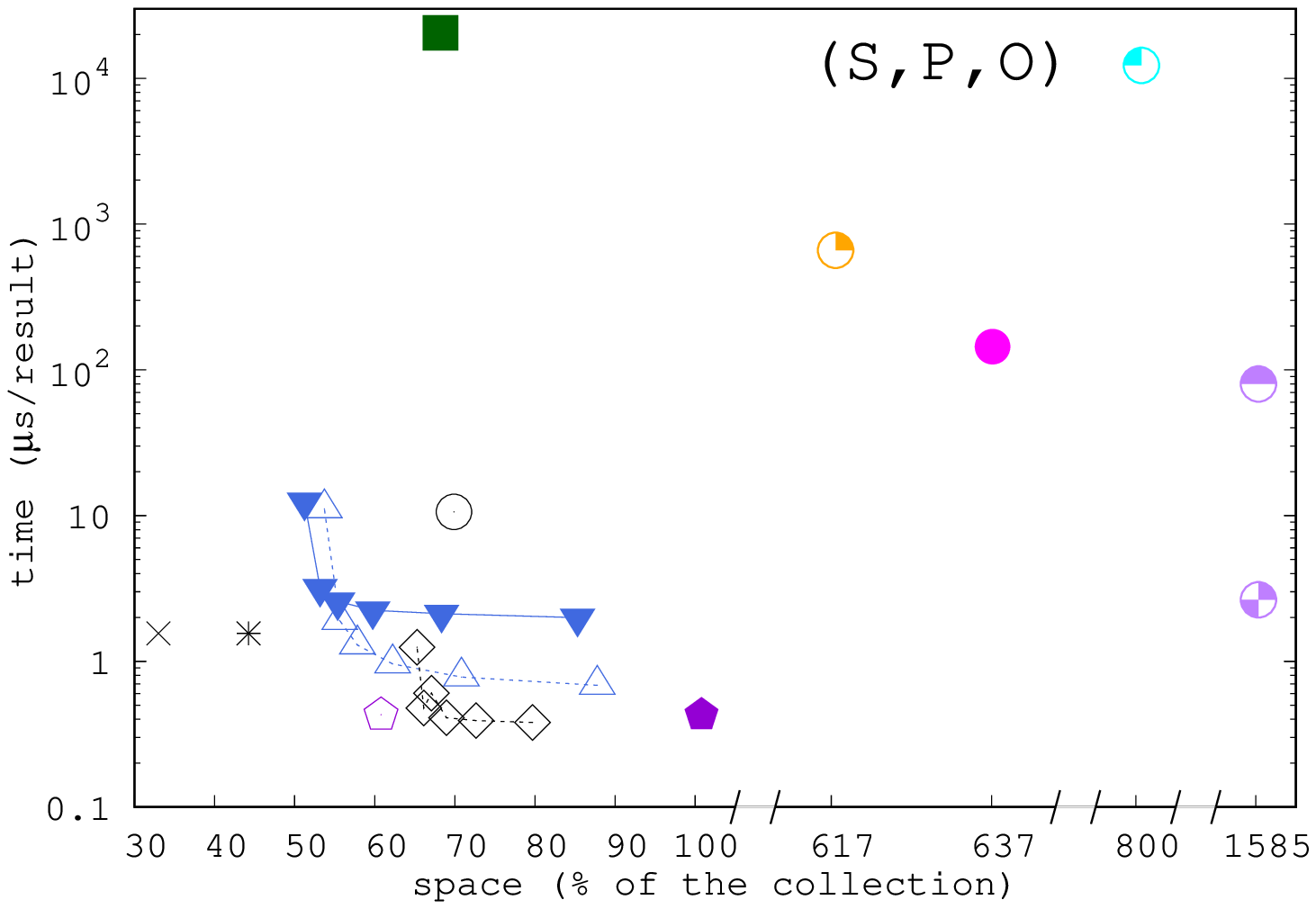}
		\includegraphics[angle=-90,width=0.33\textwidth]{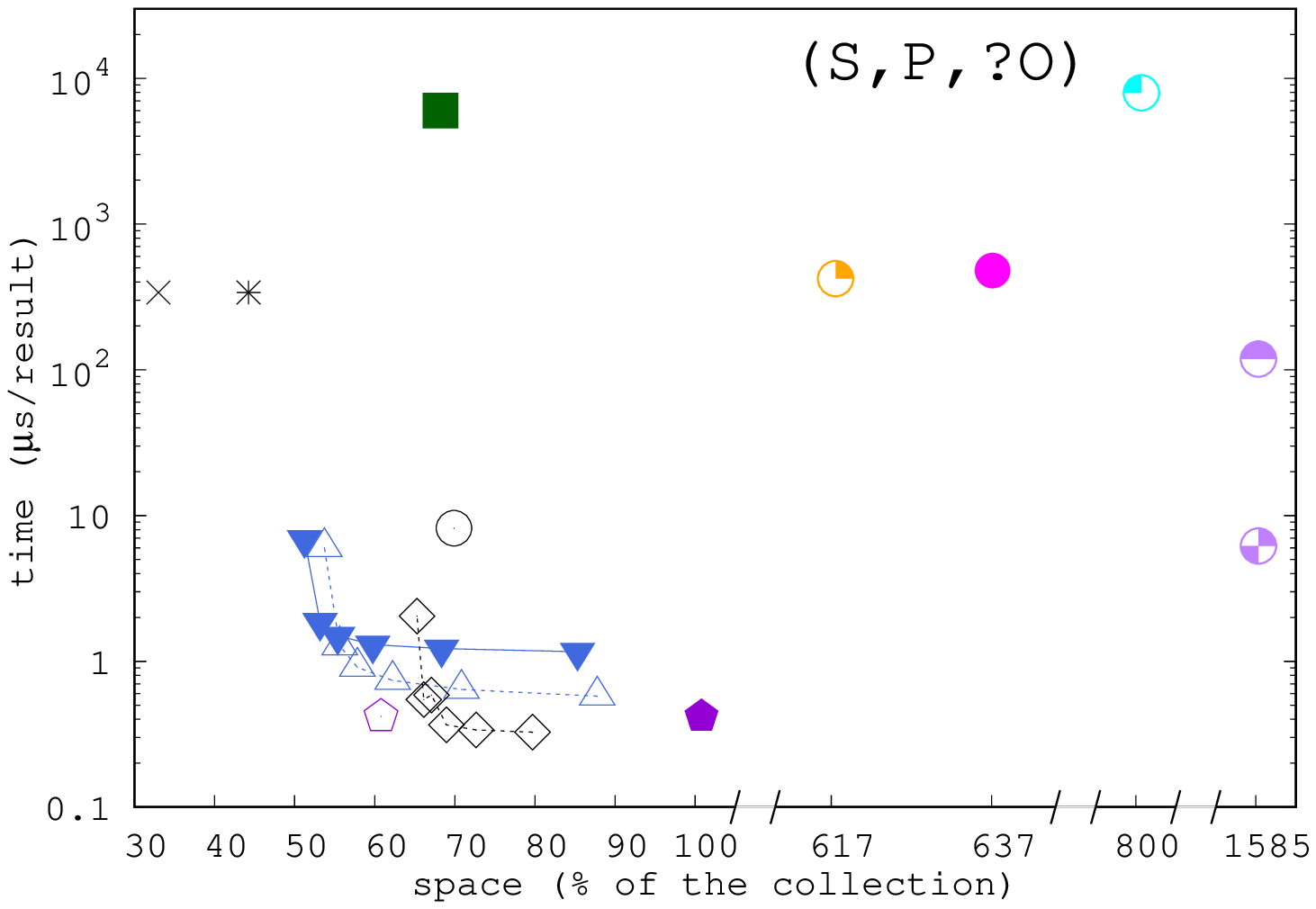}
		\includegraphics[angle=-90,width=0.33\textwidth]{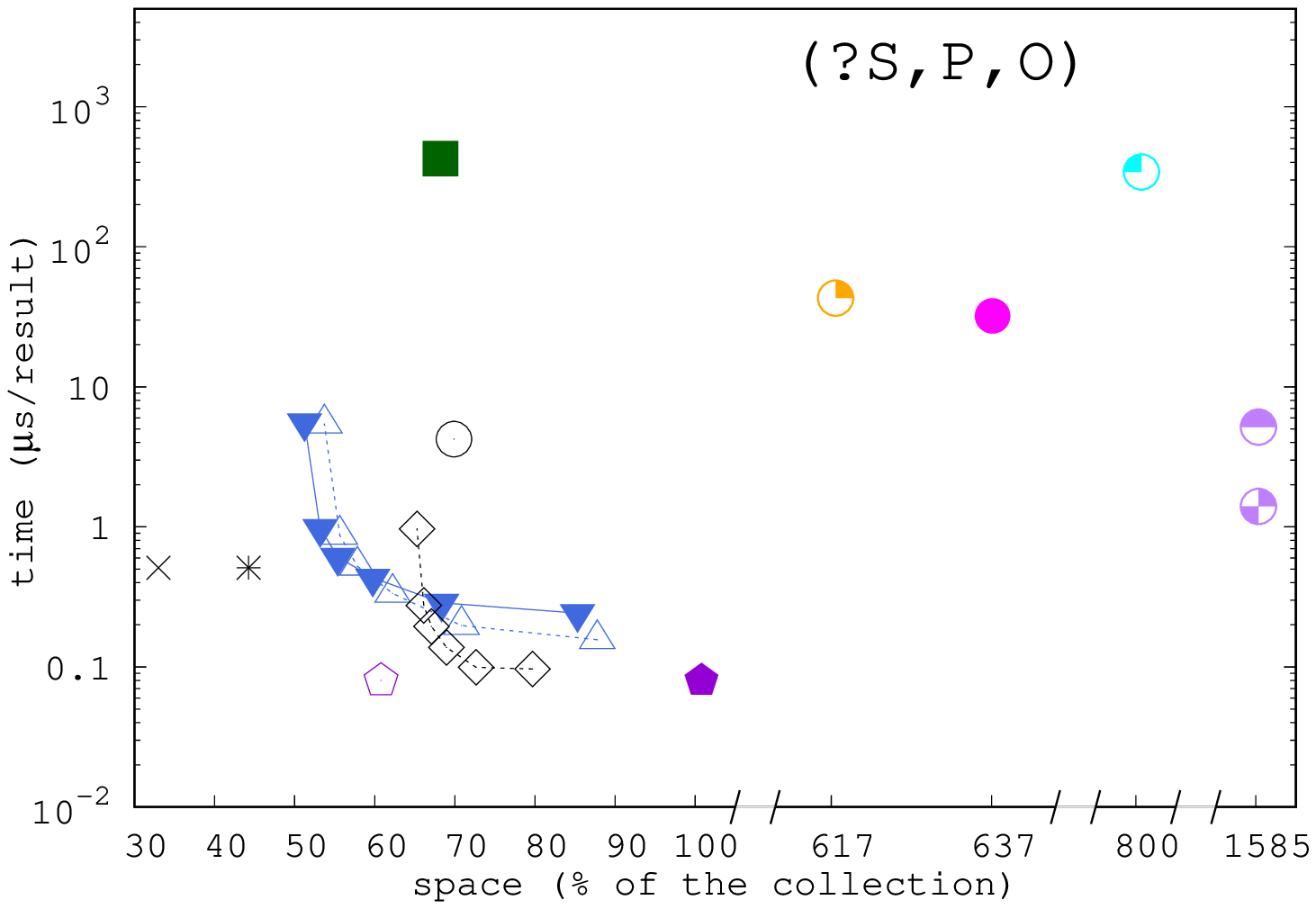}
		\includegraphics[angle=-90,width=0.33\textwidth]{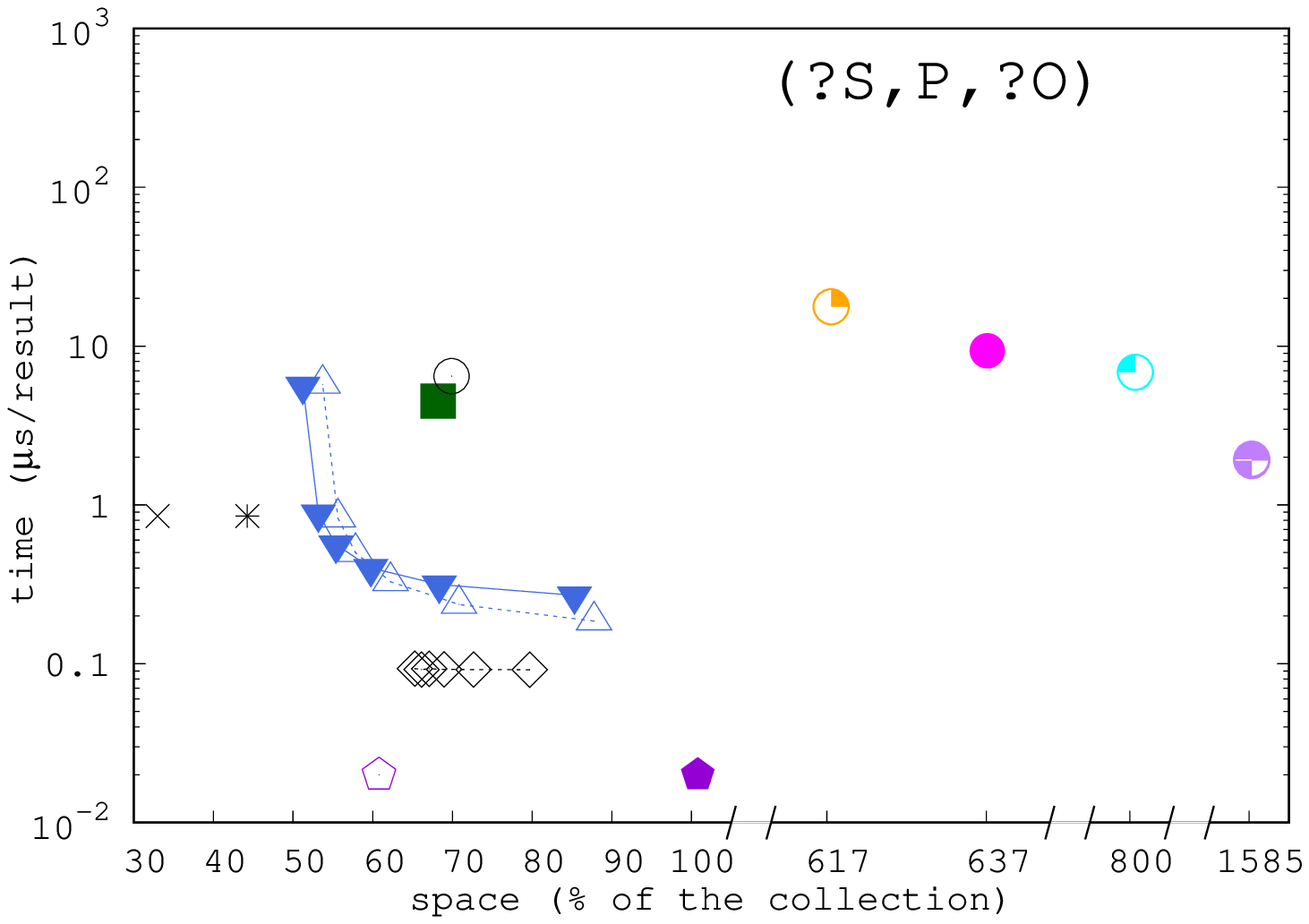}
		\includegraphics[angle=-90,width=0.33\textwidth]{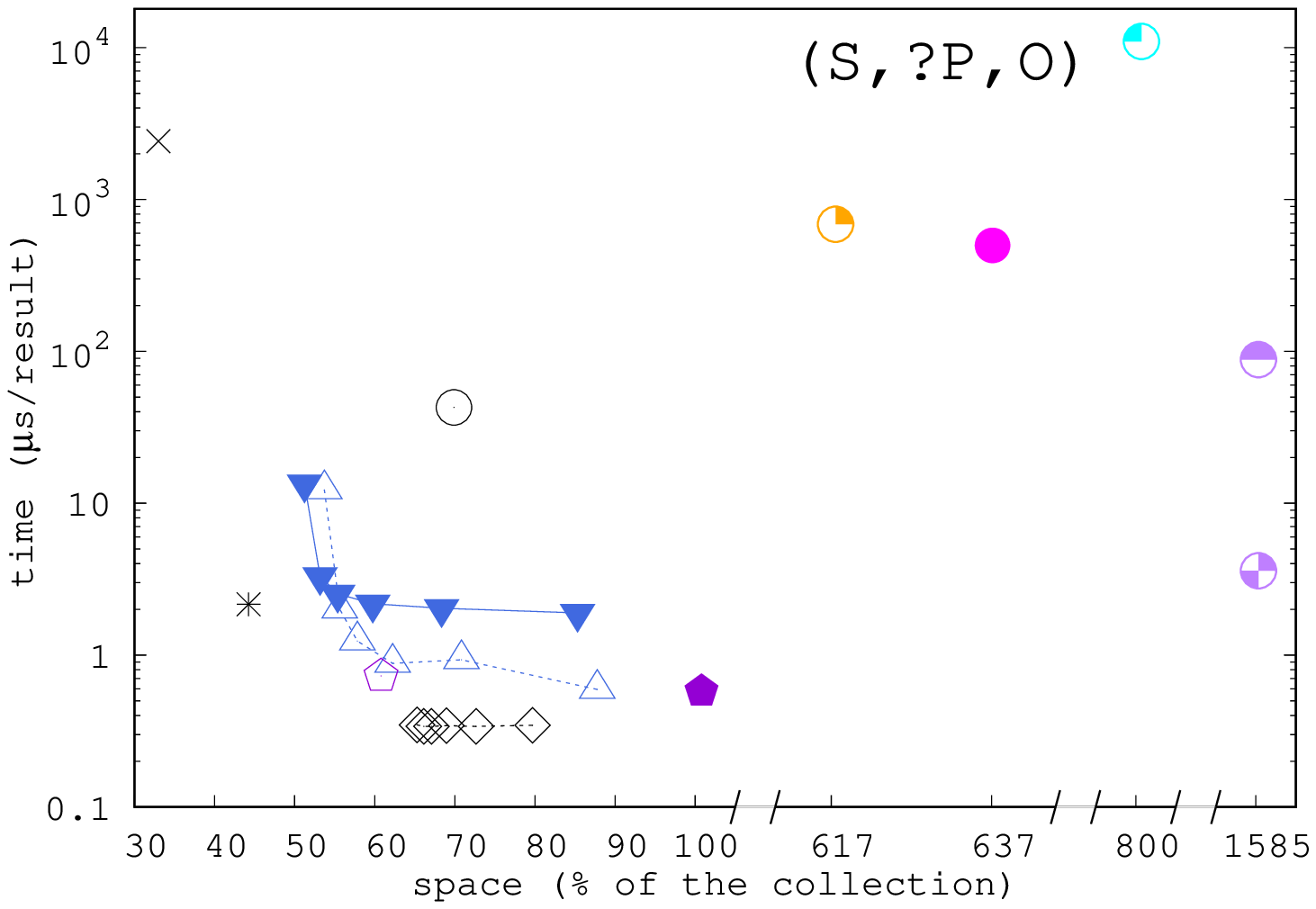}
		\includegraphics[angle=-90,width=0.33\textwidth]{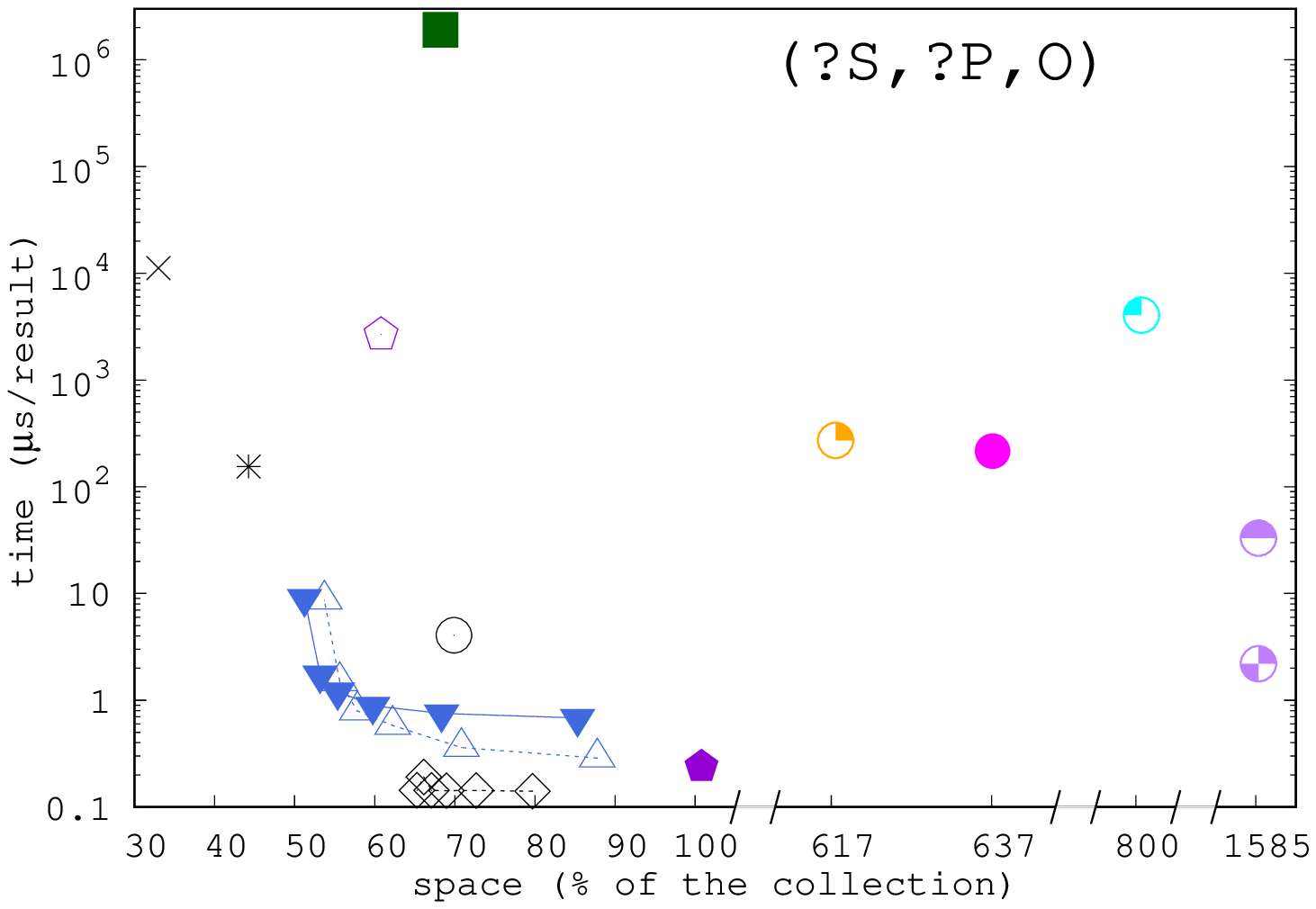}
		\includegraphics[angle=-90,width=0.33\textwidth]{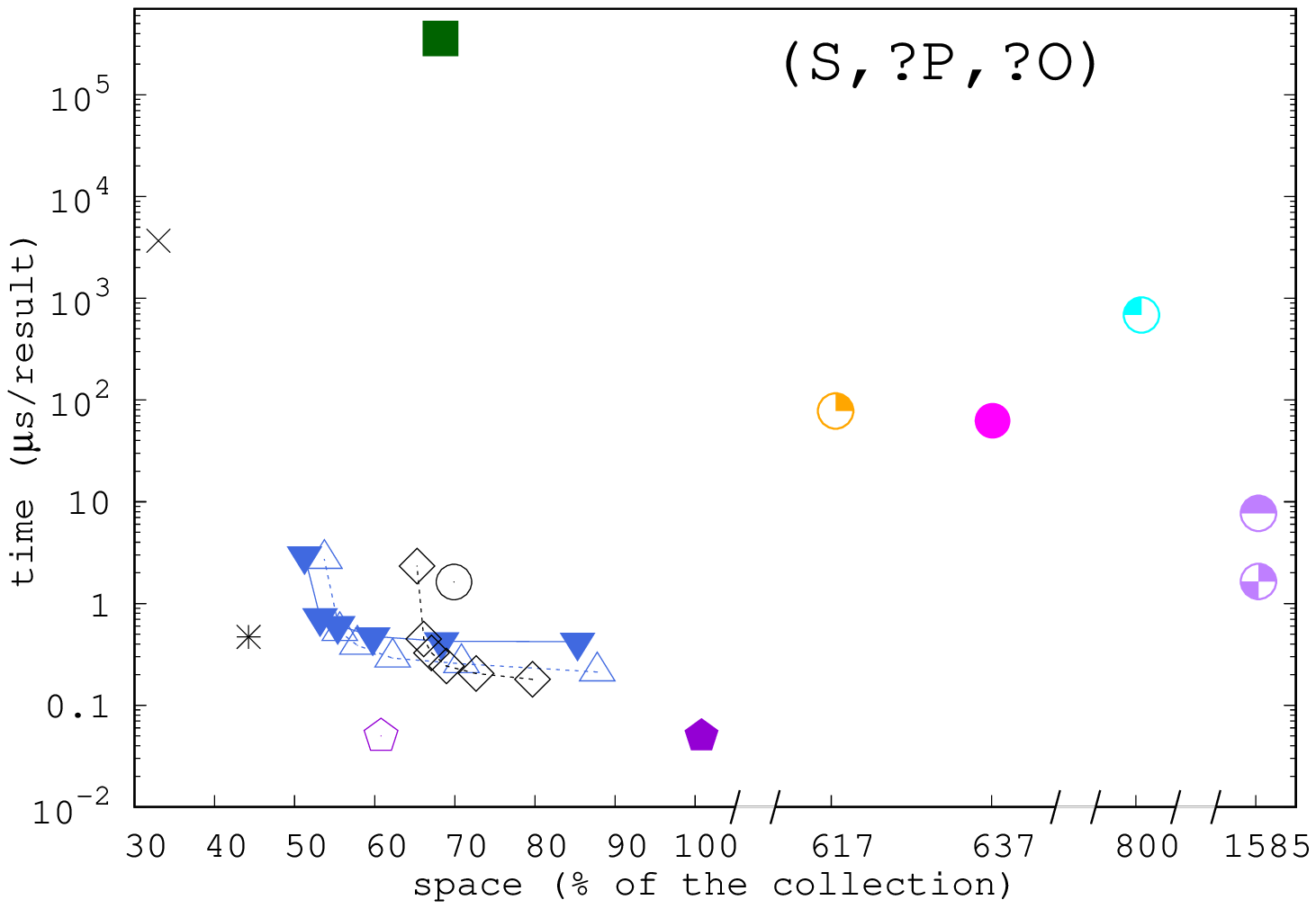}
		\includegraphics[angle=-90,width=0.33\textwidth]{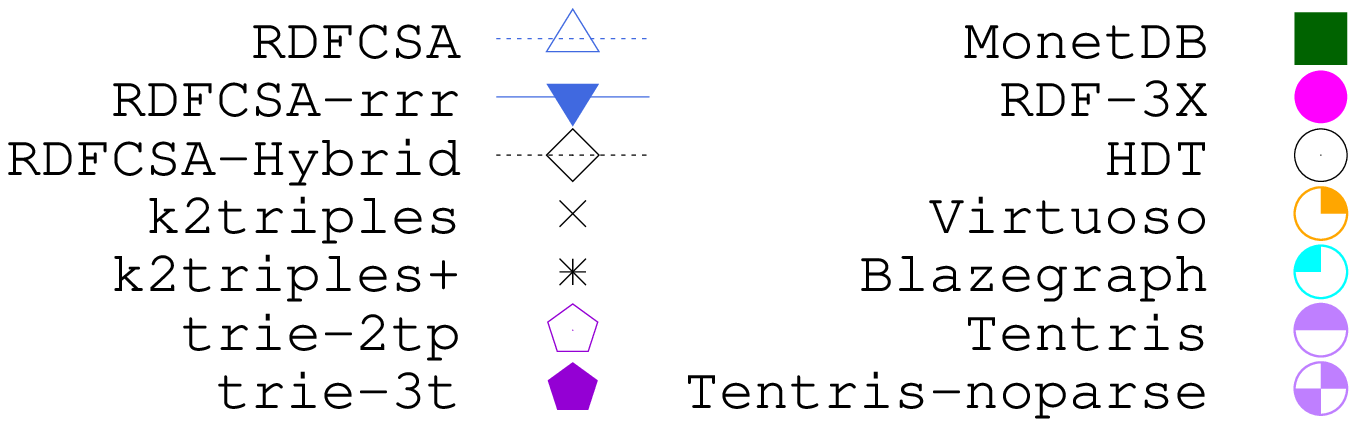}
		\caption{Space/time tradeoff on the triple-pattern queries. Note the log
			scale in the y axis. The x axis uses linear scale but includes three breaks to include all results. Query times
			in microseconds per result reported. MonetDB did not completed in reasonable time for $(s,?p,o)$ triple patterns.}
		\label{fig:experiments}
	\end{center}
\end{figure*}

Figure~\ref{fig:experiments} shows the space/time tradeoffs obtained by all the
solutions in the core triple-pattern queries. We display a plot
per triple pattern, including the values for each alternative.

Let us first focus in the comparison among our \rdfcsa variants. The
\emph{RDFCSA-rrr} variant, which aims at reducing the space of \rdfcsa, is moderately successful in that sense, with little impact in the time when the structures use little space (i.e., nearly 50\% of space thanks to a sparse sampling of $\Psi$). Thus, it is an interesting
alternative to reduce space. However, when we aim at improving the query performance by using a denser sampling of $\Psi$, the \emph{RDFCSA-rrr} becomes much slower than the basic \rdfcsa. The \emph{RDFCSA-Hybrid} variant, instead, uses at least 65\% of space, but it is significantly faster than the basic \rdfcsa. 
This variant improves its times with a denser sampling of $\Psi$ only in query patterns where the subarray $\Psi_p$ is involved.

We next focus on the comparison with other solutions. The results show that
\rdfcsa requires more space than \kdostriples, and even than the faster
\kdostriplesplus. The trie-based solutions achieve significantly different
compression rates: \trietwo is comparable in space to \rdfcsa, whereas \trieb is
up to 60\% larger. 
MonetDB and HDT are also close to the compression ratio of \rdfcsa, whereas the remaining alternatives require significantly more space: Virtuoso and RDF-3X require 7--8 times the space of \rdfcsa, Blazegraph is 10 times larger and Tentris is 20 times larger (note that a triple break is added to the x axis to display all results together, distorting the huge differences in space between these techniques). 

In addition to being much larger, Virtuoso, Blazegraph, and RDF-3X are much slower in general than the alternatives based on compact data structures. Note, however, that query parsing time is included in the measurements for these tools. In the case of Tentris, we display query times both including parsing time and excluding it, as this information is segregated by the query tool. Results show that parsing time causes a significant overhead in these queries, and ignoring this parsing time makes Tentris competitive in query times with our solutions, although using much more memory. Among the more compact solutions, the hybrid \rdfcsa yields the fastest query times in most patterns, improving on the performance of \kdostriples and achieving query times competitive with permuted trie indexes: \rdfcsa is competitive with \trieb, requiring less space, and is  more consistent than \trietwo. HDT is easily dominated by \rdfcsa variants in all query patterns.

Recall that we display the space and query times required to store and query triples of integers for the approaches based on compact data structures, but RDF-3X, Virtuoso, Blazegraph, and Tentris process the original RDF data. Space results are therefore not directly comparable, but these techniques are still a relevant baseline as SPARQL query tools. Note that \rdfcsa, \kdostriples, and permuted trie indexes could be complemented with a compact string dictionary that follows the encoding proposed for HDT. Solutions like HashDAC-RP~\cite{migueldiccionarios} can answer string-to-id and id-to-string translations in a few microseconds per operation (typically requiring 1--4 microseconds per operation in URI and literal dictionaries such as those required in DBpedia~\cite{migueldiccionarios,diccionariosCIKM}). This dictionary would increase the size of the structure by an extra 60\% of the collections in our plots, keeping them in roughly 90--150\% of the original collection (still 4 times smaller than Virtuoso, the most compact of the alternatives). 
This means that, even adding the space required for such a dictionary, \rdfcsa would still easily overcome Virtuoso, RDF-3X, Blazegraph, and Tentris in space. Additionally, since each triple-pattern query requires at most 3 string-to-id translations per query, and at most 3 id-to-string operations per returned result (at most 2 translations in practice, ignoring the $(?s,?p,?o)$ triple pattern), query times would be increased by less than 10$\mu$s per result in most cases when adding this dictionary. Note, however, that query times for Tentris ignoring parsing time (\texttt{Tentris-noparse}) are also below this limit, making it competitive in practice with \rdfcsa. Virtuoso and Blazegraph are probably affected in similar amounts by parsing overheads in these queries, making them look less competitive than they could be in practice. In Section~\ref{sec:exp:joins} we will show results for the more complex join operations, where the effect of the dictionary and query parsing overheads is less significant in general, and query times comparisons will be fairer.

\medskip
We now discuss specific results for each triple pattern, though
overall trends can be easily detected: \kdostriples and \kdostriplesplus
are the most space-efficient solutions, but their performance is
difficult to assess, since it varies significantly among triple
patterns. In turn, \rdfcsa obtains consistently low query
times, never exceeding 10 microseconds per result in any triple pattern for reasonable
sampling intervals. {\em Trie-2tp} obtains compression comparable with that of \rdfcsa
and better query times in most triple patterns, yet as explained before it has a
major drawback: the $(?s,?p,o)$ pattern is up to 10,000 times slower than
the others, and roughly 1000 times slower than \rdfcsa, effectively limiting the
application of this solution. The strongest counterpart, \trieb, on the other
hand, achieves the best query times in some cases, yet at the cost of much
worse compression ({\em RDFCSA-Hybrid} outperforms it in the others, using less space). HDT is consistent in query times, but slower and larger in general than
\rdfcsa. MonetDB is several orders of magnitude slower than \rdfcsa, using
similar space, whereas Virtuoso, Blazegraph, RDF-3X, and Tentris are much larger than our technique. Query times for Virtuoso, Blazegraph, and RDF-3X are still much higher than those of \rdfcsa in general. Results for Tentris, however, show that for most of the triple patterns the cost of query parsing is much larger than the query execution itself, that only requires a few microseconds per result. This is comparable to \rdfcsa, that would still have to be augmented with a dictionary to transform integer IDs in the result to the original strings. Nevertheless, Tentris requires over 20 times the RAM of \rdfcsa (even augmenting \rdfcsa with the string dictionary, Tentris would still be 10 times larger), so we do not consider it to be a fair competitor for \rdfcsa and the other compact solutions. 

Therefore, in what follows we focus on the comparison between \rdfcsa, \kdostriples, and \trie variants. We will resume the comparison with the remaining triple stores when testing join queries, in which the relative overhead of query parsing should be much smaller, and solutions like Virtuoso and Blazegraph become more competitive.

The simplest triple pattern, $(s,p,o)$, is the best case for \kdostriples, since it performs a single-cell retrieval query at $(s,o)$ in the \K associated with predicate $p$. In terms of time per result, this query is the worst for \rdfcsa, since it
searches for a pattern of length 3 to return at most one
occurrence. Still, \rdfcsa outperforms \kdostriples with a reasonable sampling for $\Psi$ (i.e., using over 55\% space). The variant {\em RDFCSA-Hybrid} is the fastest, together with the \trie variants.
The situation is very similar for the triple pattern $(?s,p,o)$, where \kdostriples has to scan a short column for fixed coordinate $o$ in the grid.

\kdostriples worsens by orders of magnitude in triple patterns $(s,p,?o)$, because it has to scan all the objects in a long row (fixed $s$ coordinate) of the \K associated with predicate $p$. Instead, \rdfcsa and \trie variants are almost unchanged. In fact, {\em RDFCSA-Hybrid} becomes slightly faster than the \trie variants when using 70\% space. 

In the triple pattern $(?s, p, ?o)$, \kdostriples simply retrieves all the points in the \K of predicate $p$, so its time per result is good (but still outperformed by \rdfcsa). This time, the \trie variants sharply outperform our fastest variant, \emph{RDFCSA-Hybrid}.

The lower half of Figure~\ref{fig:experiments} displays the three triple patterns where the predicate is unbound. In these patterns, \kdostriples is
very inefficient, so we compare with \kdostriplesplus,
which uses significantly more space (yet still less than \rdfcsa). As before, even the basic \rdfcsa outperforms \kdostriplesplus once using over 55\% of space, by orders of magnitude on $(?s,?p,o)$. Our fastest variant, \emph{RDFCSA-Hybrid}, also outperforms the \trie variants, except on $(s,?p,?o)$, where the latter are clearly faster. Note that the main drawback of \trietwo shows on $(?s,?p,o)$, where it is several orders of magnitude slower.

Overall, the results show that \rdfcsa is an intermediate spot between
\kdostriples, which achieves by far the best compression among the tested
solutions (but is outperformed in time by \rdfcsa), and \trieb, which disputes the best query times with our variant {\em RDFCSA-Hybrid} (but uses more space). \rdfcsa stands out as a very relevant space/time tradeoff, while offering stable and predictable times across all triple-pattern queries.
This consistency is particularly significant taking into account
that triple patterns are the basis for more complex SPARQL queries, which
perform joins involving a number of triple patterns. An inefficiency in one
triple pattern may sharply degrade the performance of the whole complex query.
This is a problem in variants like \trietwo and \kdostriplesplus, which are several orders of magnitudes slower on some triple patterns, and makes them less appealing for a general-purpose SPARQL query engine.

\subsection{Join queries}
\label{sec:exp:joins}

After analyzing \rdfcsa on basic triple patterns, we study the performance of
the different solutions in join queries involving two triple patterns. In this
section we only display results for some of the relevant state-of-the-art
alternatives used previously. Particularly, we keep \kdostriples and \kdostriplesplus, MonetDB, RDF-3X, Virtuoso, and Blazegraph. We exclude from this comparison HDT and permuted trie indexes, that have no specific mechanisms for joins, and implementing merging or chaining evaluation on top of their triple pattern queries would yield the same relative performance with respect to \rdfcsa we observed in Figure~\ref{fig:experiments}. We also omit results for Tentris, since parsing errors were returned for most of the join queries in our query sets.
Further,
for simplicity we only display results for the basic implementation (\emph{RDFCSA}) and the \emph{Hybrid} version (\emph{RDFCSA-Hybrid}). Finally, even though \rdfcsa can still obtain space/time tradeoffs for join queries, for the sake of clarity we focus the
analysis in this section on query times, and display results only for one
sampling period of $\Psi$ ($t_{\Psi}=32$, the third point left-to-right in Figure~\ref{fig:experiments}).

\begin{figure*}[tbp]
	\centering
	\includegraphics[width=0.7\linewidth]{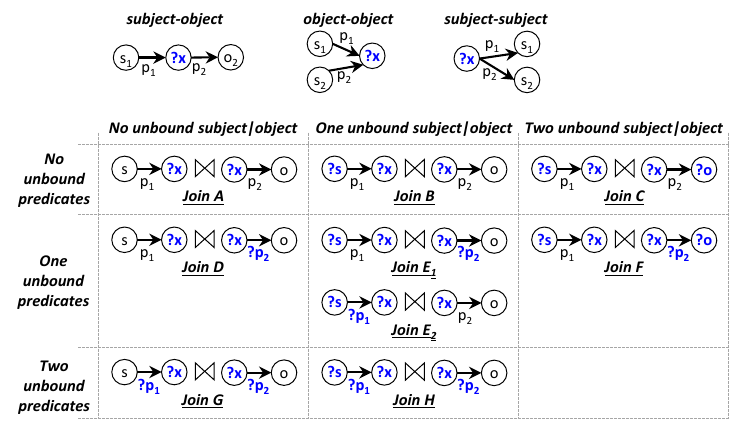}
	\caption{Join variants included in the testbed. Joins are classified 
		by number of unbound predicates, and number of unbound variables.}
\label{fig:join-types}
\end{figure*}


We analyze the results for all the different binary join queries that can
arise in practice, involving two triples, using an existing testbed~\cite{AGBFMPNkais14}. Figure~\ref{fig:join-types} displays the different query types included in the testbed and their characteristics. This testbed categorizes the joins by the number of unbound predicates, and the number of unbound subject/objects. For instance, join A has no unbound predicates, and no  unbound subject/object. Therefore, a pattern for this join is $(s,p_1,?x)\bowtie(?x,p_2,o)$. This is the subject-object variant of this join, since the join variable $?x$ is subject in one triple and object in the other. Two other variants of the same join can be created: $(?x,p_1,o_1)\bowtie(?x,p_2,o_2)$ (subject-subject join), and $(s_1,p_1,?x)\bowtie(s_2,p_2,?x)$ (object-object join). Figure~\ref{fig:join-types} details the specific bindings for subject-object joins, but the remaining configurations can be easily inferred.

In the following sections we will display results categorized according to the number of unbound predicates in the join patterns. This has little effect on performance for \rdfcsa, but severely affects tools based on vertical partitioning like MonetDB and \kdostriples (although the \kdostriplesplus variant of \kdostriples mitigates this problem with its extra indexes). In each category, queries are listed in order of increasing  ``complexity'', in the sense that additional unbound variables generally lead to a larger number of intermediate results, and therefore additional computation is
required. For instance, joins A, B, and C have no unbound predicates, and have 0, 1, and 2 unbound subject/objects respectively, so join C should be more complex in general than join A. 

The different configurations yield 9 join patterns (A, B, C, D, E.1, E.2, F, G, H), each with 3 variants: subject-subject (SS), subject-object (SO), object-object (OO). Following the original testbed, for each join type and variant we use two different query sets (\textsf{-big} and \textsf{-small}), which differ in the average number of results returned by the queries. This yields a total of 54 query sets.

Finally, for each join type, we display query times for the different join strategies applied in each case: merge-join (\textsf{-merge}), and left-
(\textsf{-left}) and right-chaining (\textsf{-right}), as well as
interactive evaluation in \kdostriples (\textsf{-int})~\cite{AGBFMPNkais14}.
Note that in some joins, specific strategies are inherently less efficient; we display all of them for \rdfcsa in our results for completeness, excluding only the alternatives that would cause a
full database query ($?s,?p,?o$). Because of the inherent inefficiency of some
techniques depending on the type of join, we will focus our discussion mainly on
the most efficient strategies for each join type. 
Moreover, for some query 
patterns and configurations we were not able to obtain results in reasonable
time with some tools: multiple query sets could not run in MonetDB, including all variants of join G and H, due to the two unbound predicates; several query sets are also omitted for RDF-3X, Virtuoso, and Blazegraph; a few query sets also failed with \kdostriples or \kdostriplesplus. When no time could be obtained, the corresponding bar will appear empty in the plots that display the results.

As discussed before, space and time comparisons between the more compact solutions (\rdfcsa and \kdostriples) and the remaining alternatives may be affected by SPARQL query parsing and other overheads that are not considered in the former. Therefore, the  direct comparison of the results could be unfair to SPARQL-compliant stores, especially in the simpler joins that return a smaller number of results. In more complex joins, in which the number of unbound variables is large, and especially in \textsf{-big} query sets, the overhead of query parsing should be smaller and time comparisons should more accurately reflect the actual query performance. Taking all of this into consideration, we will focus most of our analysis on the
comparison between \rdfcsa and \kdostriples (or \kdostriplesplus), highlighting only particular cases where the performance of the other systems should be noted.

\subsubsection{Joins with no unbound predicates}


\begin{figure}[htbp]
\centering
\includegraphics[angle=-90,width=1.0\linewidth]{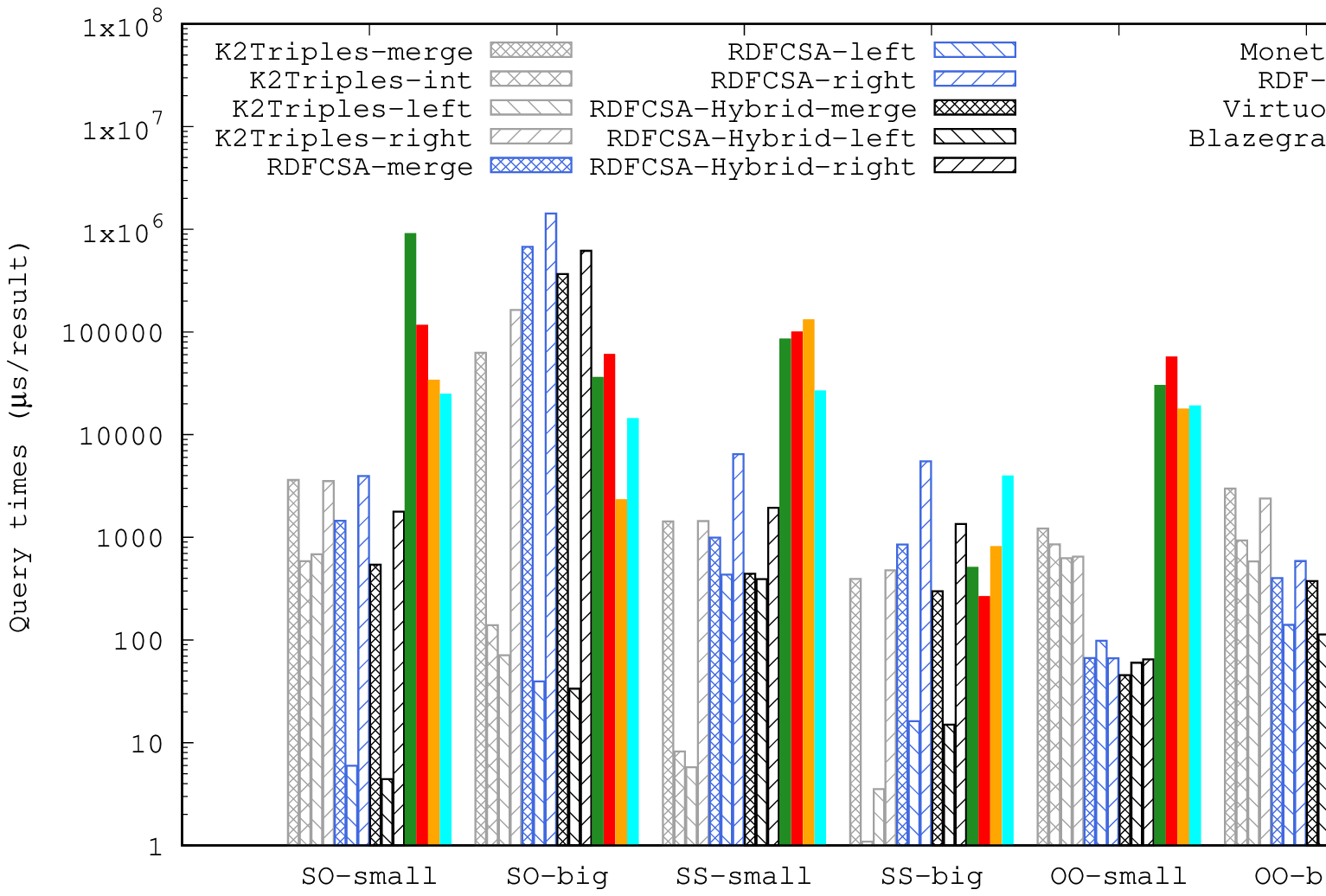}
\includegraphics[angle=-90,width=1.0\linewidth]{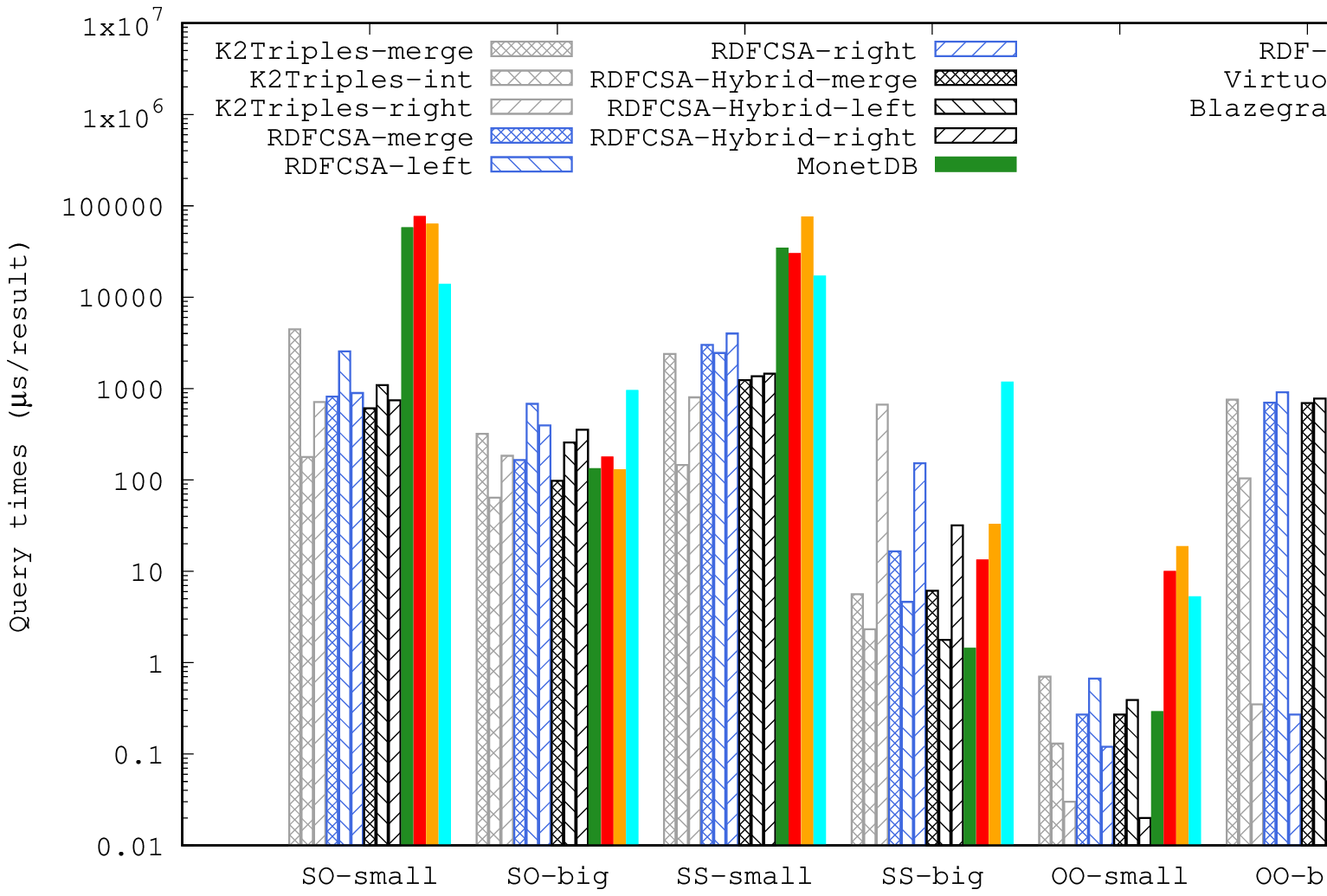}
\includegraphics[angle=-90,width=1.0\linewidth]{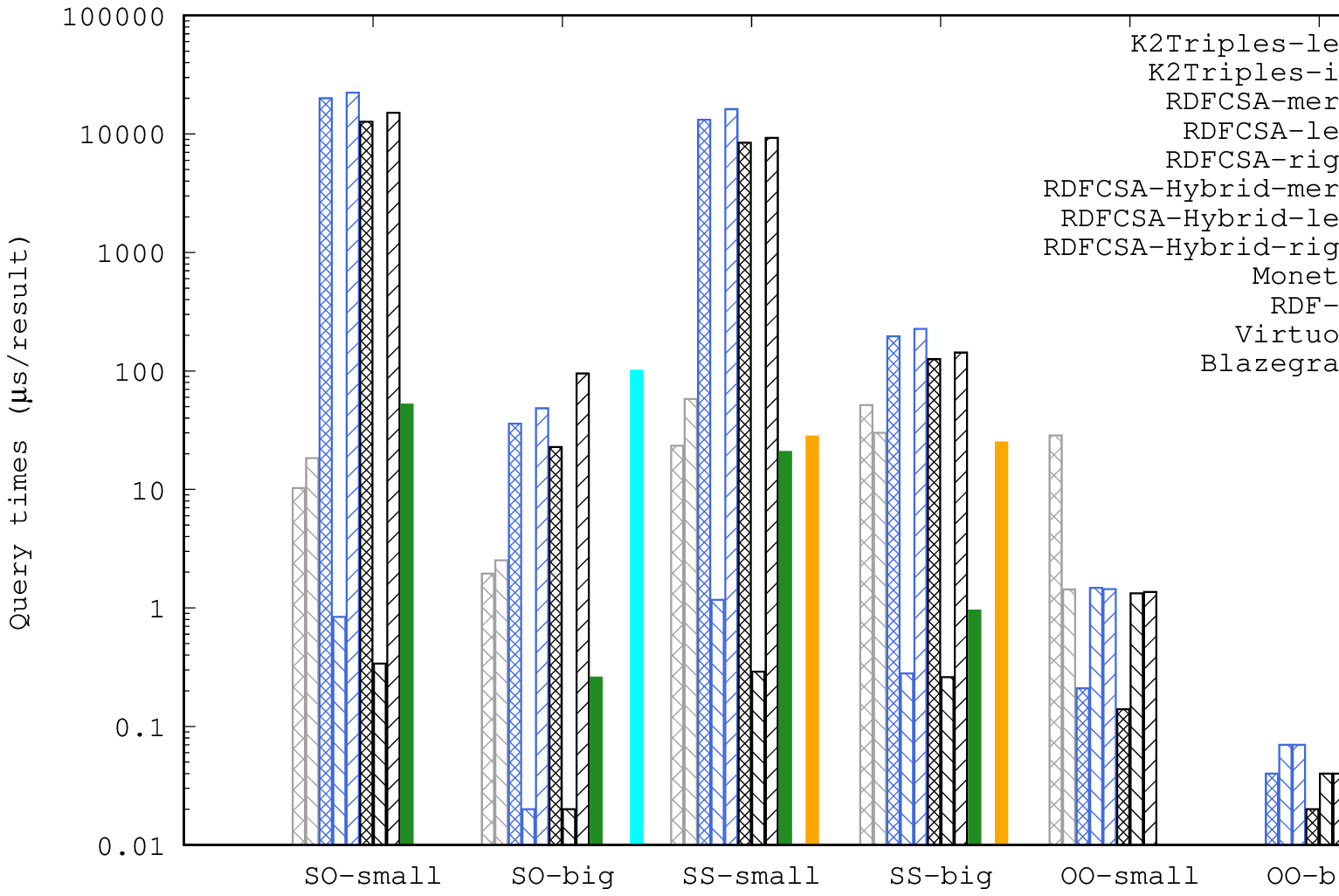}
\caption{Results for join A (top), B (middle), and C (bottom).}
\label{fig:joins147}
\end{figure}

For these joins, we display results for \kdostriples, since the additional indexes used by \kdostriplesplus do not yield any improvement in performance for fixed-predicate queries.
The top plot of Figure~\ref{fig:joins147} displays the results for
join A (e.g., $(s,p_1,?x)\bowtie(?x,p_2,o)$, with no unbound variables). In this
join, \rdfcsa with left chaining obtains the best results in all cases. This technique, for subject-object joins,
essentially executes each join as an $(s,p_1,?x)$ query chained with $(x_i,p_2,o)$
queries for each $x_i$ that results from the first query.
The results are similar for object-object joins,
but for subject-subject joins, \kdostriples obtains better query times.
This difference, depending on the position of the join variable, is consistent
with our previous results on triple patterns: when executing an
object-object or subject-object join with left chaining, the first query
executed involves an $(s,p,?o)$ pattern, where \rdfcsa was two orders of
magnitude faster than \kdostriples. However, on subject-subject joins, the first
query is an $(?s,p,o)$ pattern, where query times were similar.
MonetDB, RDF-3X, Virtuoso, and Blazegraph are typically much slower than the best variant of \rdfcsa by at least one order of magnitude. Note, however, that query parsing and other fixed overheads in these tools may be especially significant in these joins, that return a very small set of results.

The middle plot of Figure~\ref{fig:joins147} displays results for
join B (e.g., $(?s,p_1,?x)\bowtie(?x,p_2,o)$, with one unbound variable). Several times, \kdostriples obtains the best query times with its interactive evaluation strategy, but {\em RDFCSA-Hybrid} is the best in the other cases. The nature
of this join, where one pattern has an extra unbound variable, leads to
uncertainty in the complexity of the best operation order. Because of this, the
interactive evaluation in \kdostriples is a good approach, even though differences are usually small. 
MonetDB, RDF-3X, Virtuoso, and Blazegraph  are competitive in some cases, especially in the \textsf{-big} executions in which the query parsing overhead is reduced and their ability to extract larger results sets is highlighted.

The bottom plot of Figure~\ref{fig:joins147} displays results for
join C (e.g., $(?s,p_1,?x)\bowtie(?x,p_2,?o)$, with two unbound variables). In
this type of join, \rdfcsa again obtains the best query times, usually with
left-chaining evaluation. This is clearly the
most efficient technique for this join, with results similar to those of join A.
When both triple patterns have a similar structure (i.e., the same number of
fixed and bound variables), \rdfcsa tends to be more efficient with
left-chaining, due to the performance of the triple-pattern queries that are generated: 
in subject-object joins, with left-chaining, we run an $(?s,p_1,?o_1)$
query followed by many $(s_i,p_2,?o_2)$ queries, which are very efficient in
\rdfcsa. However, in object-object joins the merge strategy is better. Regarding MonetDB, Blazegraph, and Virtuoso, we obtain similar results as for Join B (i.e. they are at least one order of magnitude slower than the best choice), yet we can see that in most cases we could not get results for those techniques.

\subsubsection{Joins with one unbound predicate}

\begin{figure}[tbp]
\centering
	\includegraphics[angle=-90,width=1.0\linewidth]{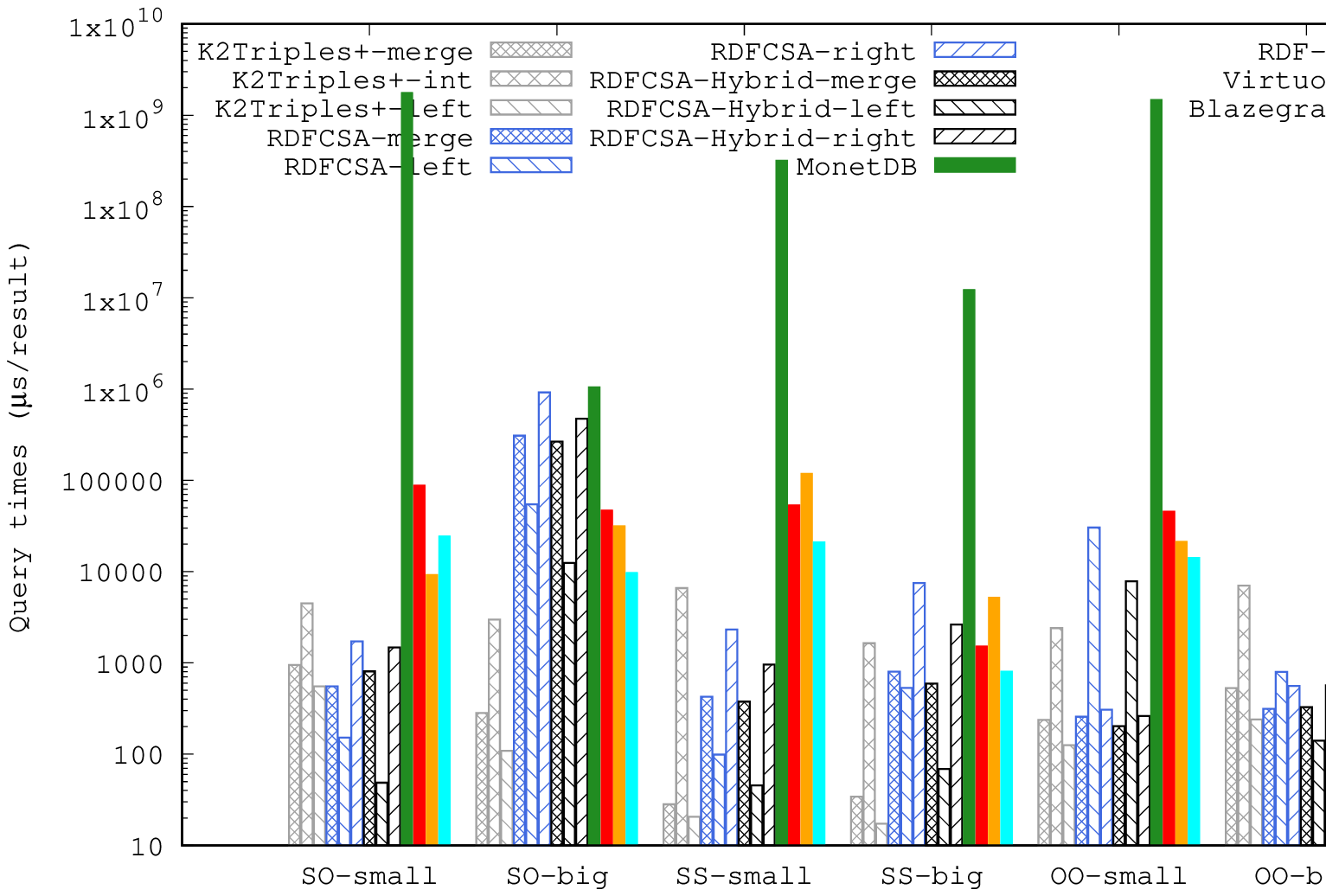}
	\includegraphics[angle=-90,width=1.0\linewidth]{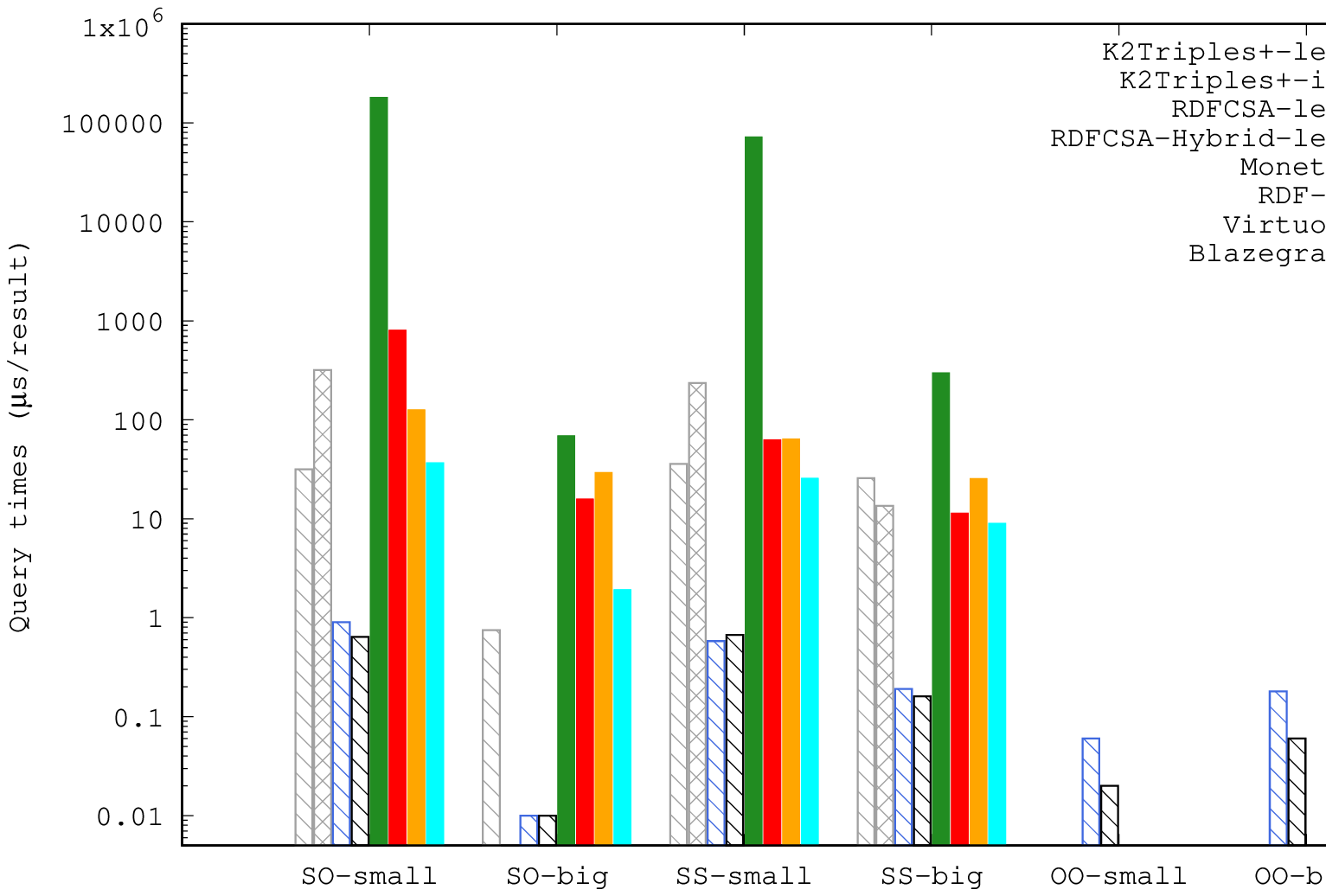}
	\caption{Results for join D (top) and F (bottom).}
	\label{fig:joins258}
\end{figure}

\begin{figure}[htbp]
\begin{center}
	\includegraphics[angle=-90,width=1.0\linewidth]{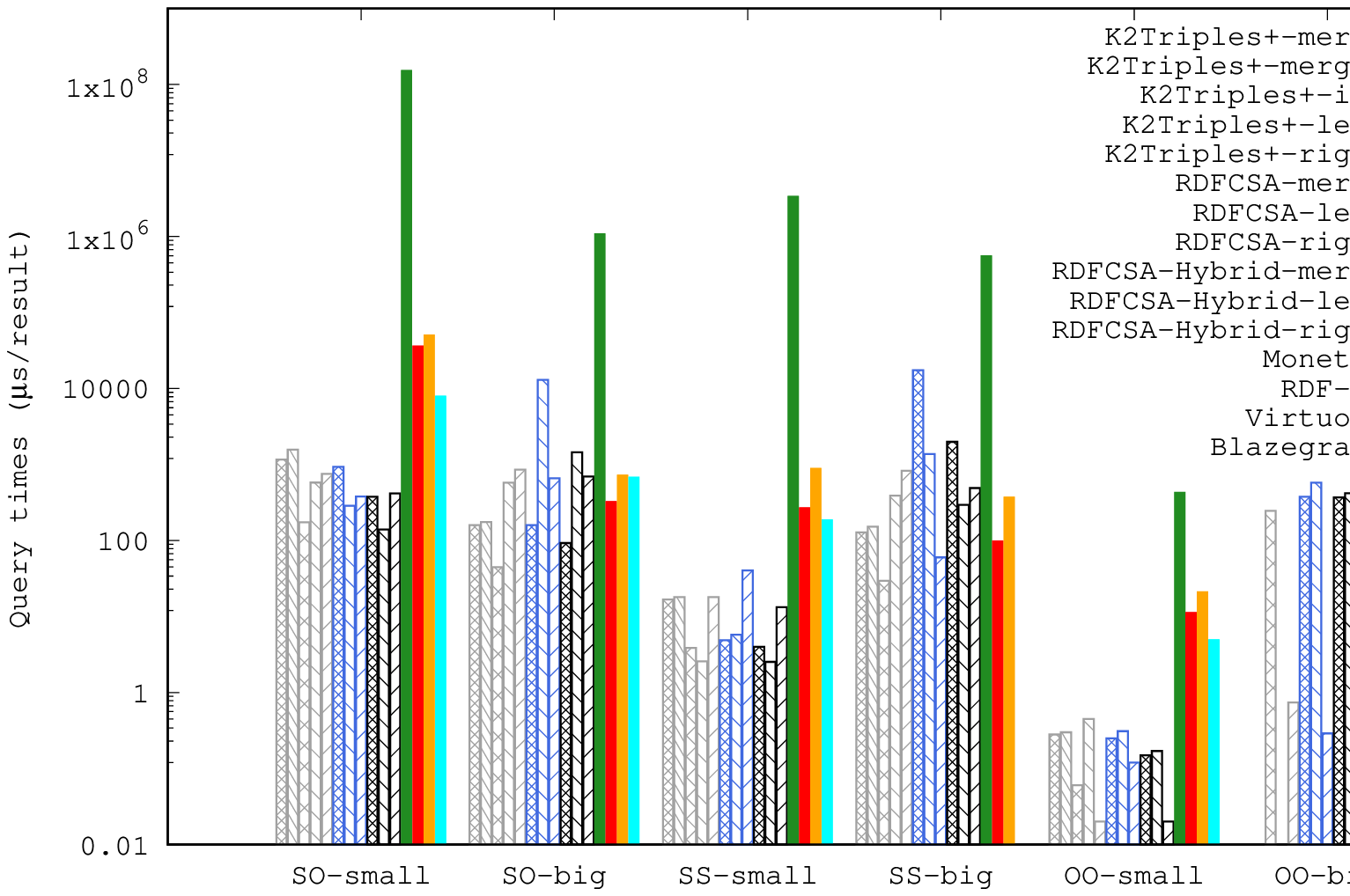}
	\includegraphics[angle=-90,width=1.0\linewidth]{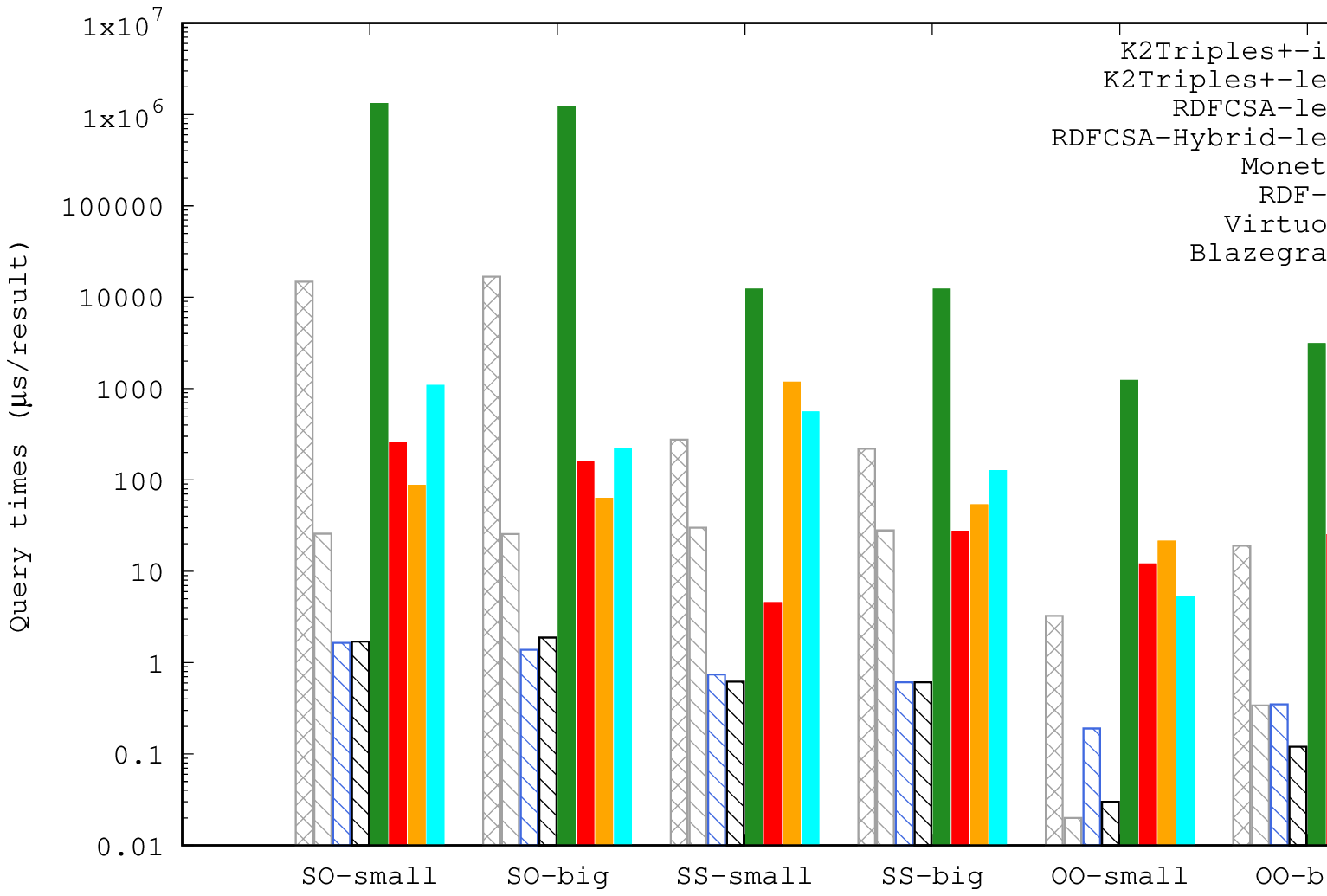}
	\caption{Results for joins E1 (top) and E2 (bottom).}
	\label{fig:join52}
\end{center}
\end{figure}

Figures~\ref{fig:joins258} and \ref{fig:join52} 
display the query times for joins D, E, and F.
In these experiments we compare \rdfcsa with \kdostriplesplus instead of
\kdostriples, since the latter is typically orders of magnitude slower. 

Considering the results across all the joins, \rdfcsa achieves better query
times. Yet, results are significantly different depending on the join type and
query set. MonetDB is far from competitive as long as an unbound predicate appears, as expected, and it is up to 5 orders of magnitude slower than the other techniques. RDF-3X, Virtuoso, and Blazegraph are also 1--2 orders of magnitude slower than the fastest \rdfcsa variant in most cases. Hence, we will focus on the comparison
between \rdfcsa and \kdostriplesplus.

In join D, \rdfcsa obtains the best overall results for object-object joins, but
\kdostriplesplus is also competitive. \kdostriplesplus is faster in subject-subject joins and in some cases for subject-object joins. Left-chaining
is the best strategy in most cases, both in \kdostriplesplus and \rdfcsa, since it evaluates the
triple pattern with bound predicate first, therefore saving a significant effort
on the right triple pattern. 

Regarding join F, \rdfcsa is
significantly faster in all cases, again with left-chaining, as this  reduces the cost of processing the pattern
with unbound predicate. Note that, for this join, most alternatives failed to yield results for the object-object joins in our setup. Finally, note that when comparing joins D and F, we find the same
trend existing between joins A and C: \kdostriples and
\kdostriplesplus are more competitive with few unbound variables. In more complex queries, instead, \rdfcsa
is much more efficient.

Figure~\ref{fig:join52} shows  significant differences between joins E1 and E2, because the different location of the unbound
predicate leads to very different triple patterns in each side of the join. The join
E1 (e.g., $(?s,p_1,?x)\bowtie(?x,?p_2,o)$) requires much more computation with any of
the basic strategies, since both triple patterns contain an unbound variable. The best evaluation strategy is unclear: the
merge and right-chaining techniques are competitive in \rdfcsa, but 
\kdostriplesplus is slightly faster in most cases with its interactive
evaluation. However, in join E2, the left pattern is much simpler than the right
one, leading to a clearer evaluation path: left-chaining is the best strategy,
and \rdfcsa is an order of magnitude faster than \kdostriplesplus in
most joins.

\subsubsection{Two unbound predicates}

\begin{figure}[htbp]
\begin{center}
	\includegraphics[angle=-90,width=1.0\linewidth]{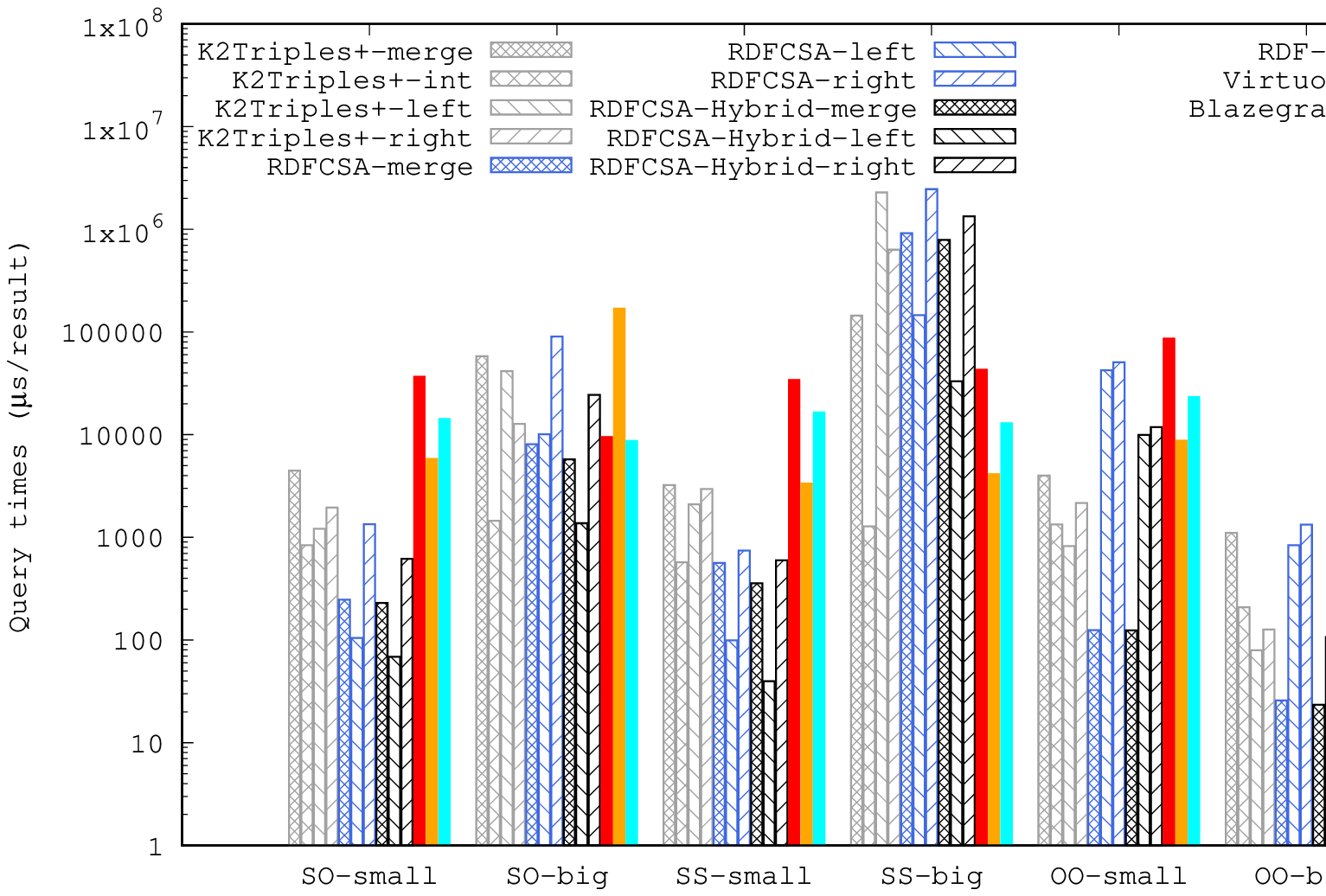}
	\includegraphics[angle=-90,width=1.0\linewidth]{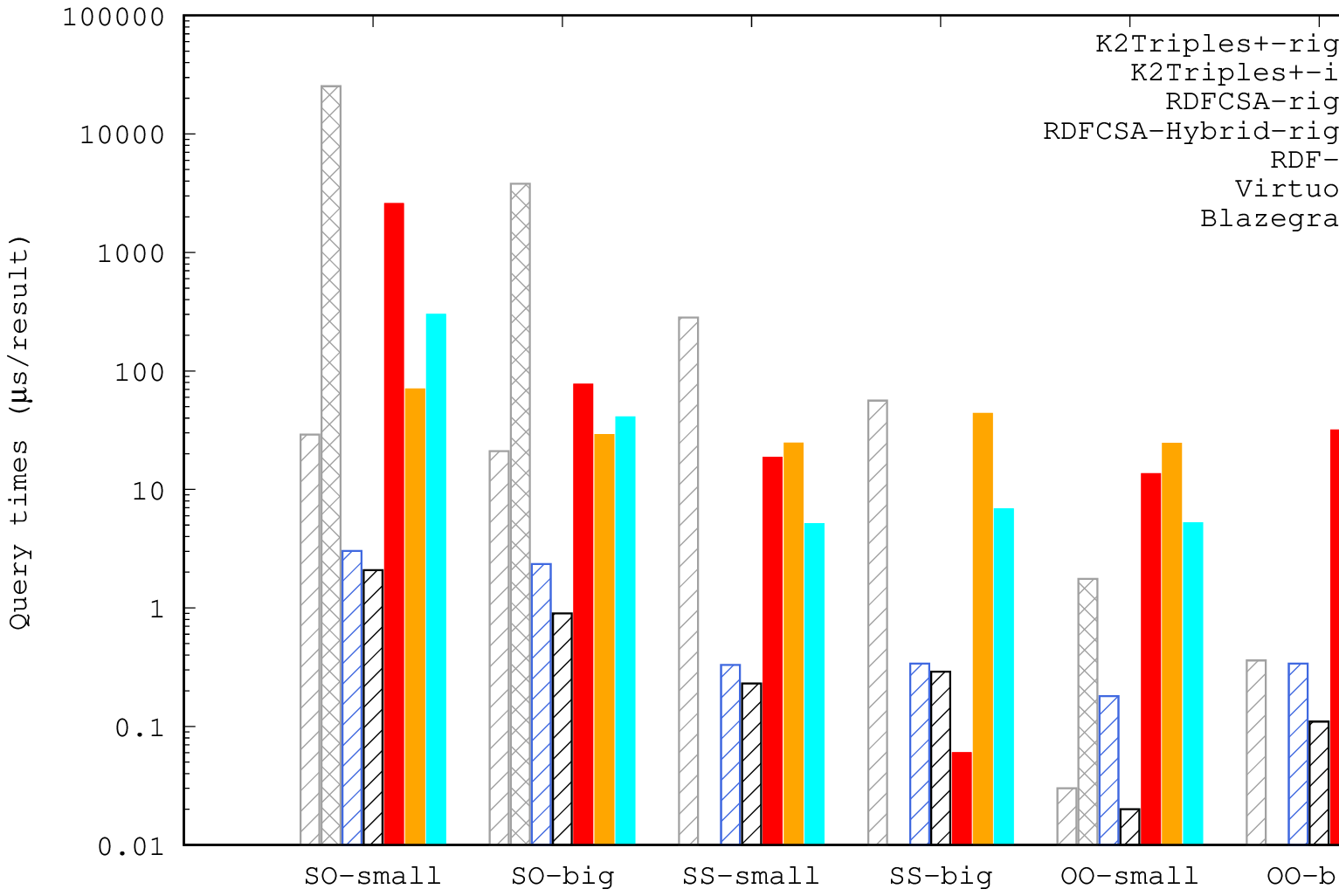}
	\caption{Results for joins G (top) and H (bottom).}
	\label{fig:join36}
\end{center}
\end{figure}

Figure~\ref{fig:join36}
displays the query times for joins G (e.g., $(s,?p_1,?x)\bowtie(?x,?p_2,o)$) and
H (e.g., $(?s,?p_1,?x)\bowtie(?x,?p_2,o)$). We omit MonetDB in these joins because the
combination of two unbound predicates makes those queries extremely inefficient in its vertical
partitioning model.

Like in previous cases, the results vary significantly depending on the join and
query set. For join G, \rdfcsa is the fastest technique in almost all cases, using merging or left-chaining depending on the case. For join H, \rdfcsa with right-chaining is
also orders of magnitude faster than \kdostriplesplus in general. These results are
again consistent with the trend in previous sections that suggests that \rdfcsa
is especially competitive in the more complex join patterns.
The subject-subject joins with many results are the only observed case where RDF-3X, Virtuoso or \kdostriples are faster than \rdfcsa. In most other query sets, however, the fastest \rdfcsa variant is 1--2 orders of magnitude faster than the other alternatives.

\section{Conclusions}
\label{sec:conclusions}

We have introduced \rdfcsa, a compact data structure for the
efficient storage and querying of RDF datasets. It is based on a compressed text index, the \csa \cite{Sad03},
which is adjusted so that the triples that compose the RDF dataset are regarded as
circular strings of length 3. We demonstrate that all the SPARQL triple patterns boil down to text searches in this particular collection of cyclic strings. The basic capabilities of \rdfcsa are then based on the
\csa search algorithms, which we have adapted and optimized for our scenario. We also design algorithms to solve queries involving joins. 

\rdfcsa is able to compress a set of RDF triples to around 60\% of their
raw size. Within this space, it offers fast and very consistent query
times for all the basic triple-pattern queries, which are the basis for SPARQL
support. In our experiments, \rdfcsa answers any triple-pattern query within a few microseconds per result. It is also able to efficiently answer queries involving binary joins, being faster in most cases than the alternatives.
Our experimental evaluation shows that state-of-the art solutions like RDF-3X, Virtuoso or Blazegraph are much larger, and in many cases slower than \rdfcsa, even considering effects such as query parsing and dictionary encoding. We also clearly outperform HDT \cite{FernaNdez:2013:HDT} in both space and time. Modern in-memory alternatives such as Tentris can achieve competitive query times with our solution, but their memory requirements are an order of magnitude higher than ours.

While \kdostriples \cite{AGBFMPNkais14} obtains better compression than \rdfcsa, its query times are much
less consistent, being several orders of magnitude slower in some triple-pattern
queries. The recent permuted trie indexes \cite{Venturini}, on the other hand, are able to outperform \rdfcsa in time, but in order to achieve consistent performance for all triple patterns
they need to use around 50\% more space. Our implementation variants also provide a wide space/time tradeoff, that can be easily tuned by adjusting the sampling interval on $\Psi$.


Overall, \rdfcsa provides a very appealing space/time tradeoff for the
storage of RDF data, combining low space with fast and consistent query times. Such predictability is very important when building up more complex SPARQL queries on top of simple triple patterns and joins. 

Our current implementation is designed to
handle integer-based triples, so it requires an external dictionary to handle
the mapping between strings and ids. As future work, we plan to integrate
\rdfcsa with some compressed dictionary \cite{migueldiccionarios,diccionariosCIKM,Martinez-Prieto:2012} in order to provide efficient mappings. Another choice is to integrate it in the HDT library (\texttt{http://rdfhdt.org}), which already provides the needed string dictionaries.
Another future challenge is to make \rdfcsa dynamic, that is, allow adding and removing triples from the database. This is already supported by indexes like RDF-3X and solutions like Virtuoso and Blazegraph; a dynamic implementation of \kdostriples also exists~\cite{k2treedinamico}. We believe that it is possible to build on dynamic variants of the \csa \cite{CHLS07,MNtalg08,MNV15} to obtain an efficient dynamic \rdfcsa.
Finally, compressed indexes inspired in the \rdfcsa\ have been used to implement multi-join algorithms in worst-case-optimal time~\cite{optimaljoins2019,optimaljoins}, which for complex queries using the same variables several times are more efficient than query plans based on binary joins.

\section{Data Availability}
The dataset DBPedia used in our experiments is available at \url{http://downloads.dbpedia.org/3.5.1/}. A processed version of the dataset, containing only the integer ids used by our representation, has been made available at \url{https://lbd.udc.es/research/rdf/}. The source code of \rdfcsa and execution scripts are also available at the same url. The testbed used is available at \url{http://dataweb.infor.uva.es/queries-k2triples.tgz}.

\ifCLASSOPTIONcompsoc
  \section*{Acknowledgments}
\else
  \section*{Acknowledgment}
\fi


Funding for the Spanish group: project PDC2021-121239-C31 (FLATCITY-POC) funded by MCIN/AEI/10.13039/501100011033,“NextGenerationEU”/PRTR”; project PDC2021-120917-C21 (SIGTRANS) funded by MCIN/AEI/10.13039/501100011033, “NextGenerationEU” /PRTR”; PID2020-114635RB-I00 (EXTRACompact) funded by MCIN/ AEI/10.13039/501100011033;  
PID2019-105221RB-C41 (MAGIST) funded by MCIN/ MCIN/ AEI/10.13039/501100011033; project RTI-2018-098309-B-C32 (BIZDEVOPSGLOBAL), funded by MICIU/FEDER-UE; grant ED431C 2021/53 (GRC) funded by GAIN/Xunta de Galicia; grant ED431G 2019/01 (CSI) funded by Xunta de Galicia, FEDER Galicia 2014-2020 80\%, SXU 20\%; also funded by Xunta de Galicia / Igape/ IG240.2020.1.185.
Gonzalo Navarro is partially funded by the Millennium Institute for Foundational Research on Data (IMFD), Chile.  
%

\bibliographystyle{IEEEtran}
\bibliography{refs}

\begin{IEEEbiography}[{\includegraphics[width=1in,height=1.25in,clip,keepaspectratio]{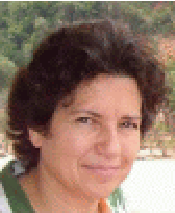}}]{Nieves R. Brisaboa}
 is the founder of the Database Laboratory of the University of A Coru\~na (\url{https://lbd.udc.es}). Since 2007, she is a full professor at the University of A Coru\~na. She has been main researcher of more than 15  national and international research projects, and led about 40 research projects with public and private institutions.  Her research interests include digital libraries, text retrieval, compressed text retrieval, deductive databases and spatial databases.
 She is author of more than 35 papers in ISI journals, and 80 papers in relevant international conferences. She has co-advised 13 PhD thesis.
\end{IEEEbiography} 

\begin{IEEEbiography}[{\includegraphics[width=1in,height=1.25in,clip,keepaspectratio]{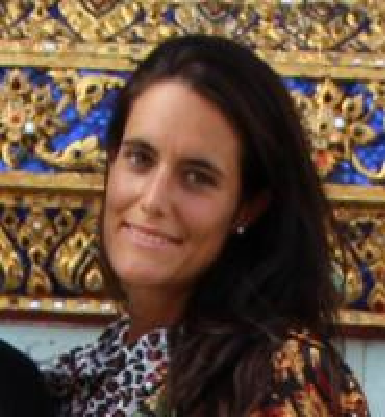}}]{Ana Cerdeira-Pena}
is an Assistant Professor at University of A Coru\~na since 2013, year when she obtained her Ph.D. degree in Computer Science from the same institution. Her fields of interest include the analysis and design of compact data structures and algorithms for data compression and indexing, mathematical modeling and algorithms design for operational research problems, and information systems management.  She has co-authored many articles in various international journals and relevant conferences, and has actively participated in several national and international research projects.
\end{IEEEbiography}

\begin{IEEEbiography}[{\includegraphics[width=1in,height=1.25in,clip,keepaspectratio]{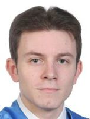}}]{Guillermo de Bernardo}
is a researcher from the Databases Lab and assistant professor at University of A Coru\~na. He received his Ph.D. in Computer Science from University of A Coru\~na in 2014. His research interests are mainly focused on data compression, and include compact data structures and algorithms, compressed text retrieval and geographic information retrieval.
\end{IEEEbiography}

\begin{IEEEbiography}[{\includegraphics[width=1in,height=1.25in,clip,keepaspectratio]{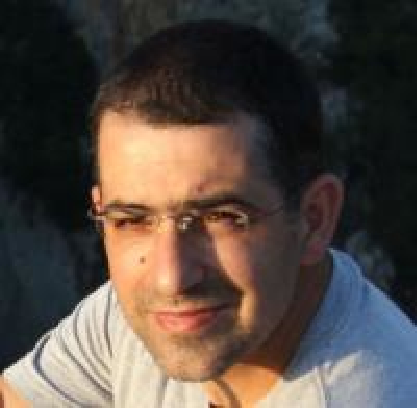}}]{Antonio Fari\~na}
is an associate professor of the Univesity of A coru\~na since 2012, where he presented his Ph.D. in 2005 in the area of text compression. His research interests include compression and self-indexing for text, graphs, and trajectories of moving objects among others. He is author of 17 ISI-Journal Papers, 24 works on relevant conferences (SIGIR, DCC, CIKM, SPIRE). He has co-advised three PhD thesis and leaded three national research projects. Apart from that, he also got involved in more than 15  research projects.
\end{IEEEbiography}

\begin{IEEEbiography}[{\includegraphics[width=1in,height=1.25in,clip,keepaspectratio]{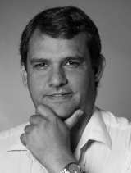}}]{Gonzalo Navarro}
is a full professor at the University of Chile. He has directed the Millennium Nucleus Center for Web Research, and leaded and participated in many research projects. In 2018, we has awarded as ``ACM Distinguished Member". He has been PC (co-)chair of conferences such as SPIRE, SCCC, SIGIR-Posters, SISAP, LATIN, and CPM. He is a member of the Editorial Board of journal such as Inf. Retrieval, ACM JEA, and Inf. Systems, and has been guest editor of special issues of ACM SIGSPATIAL, JDA, Inf. Systems, and Algorithmica. He has been PC member of more than 50 international conferences and reviewer of around 40 journals. He has given around 50 invited talks in several universities and conferences. He is author of more than 160 papers in international journals, and 240 papers in international conferences. He is author of two books published by Cambridge University Press, around 25 book chapters, editor of 10 international conference proceedings. He is one of the most prolific and highly cited authors in Latin America. 
\end{IEEEbiography}

\end{document}